\documentclass[prd,showpacs,nofootinbib,preprintnumbers]{revtex4}

\usepackage{latexsym}
\usepackage{amssymb}
\usepackage{amsmath}
\usepackage{array}
\usepackage[dvips]{epsfig}
\usepackage{amscd}
\usepackage{xspace}

\newcommand{\comment}[1]{}
\newcommand{\vm}[1]{\vphantom{#1}}
\newcommand{\hsp}{\hspace{0.15em}}
\newcommand{\nhsp}{\hspace{-0.15em}}
\newcommand{\smhsp}{\hspace{0.06em}}

\newcommand{\ds}{\displaystyle}
\newcommand{\Eps}{\mathcal{E}}
\newcommand{\lbr}{\left(}
\newcommand{\rbr}{\right)}

\newcommand{\la}{\lambda}
\newcommand{\lb}{\label}
\newcommand{\om}{\omega}
\newcommand{\al}{\alpha}

\newcommand{\vak}{\varkappa}
\newcommand{\pa}{{\partial}}

\newcommand{\si}{\sigma}
\newcommand{\Si}{\Sigma}

\newcommand{\fr}{\frac}
\newcommand{\cp}{{\cal{P}}}

\newcommand{\bw}{\begin{widetext}}
\newcommand{\ew}{\end{widetext}}
\newcommand{\be}{\begin{align}}
\newcommand{\ee}{\end{align}}
\newcommand{\ba}{\begin{eqnarray}}
\newcommand{\ea}{\end{eqnarray}}

\newcommand{\va}{\varepsilon}
\newcommand{\ep}{\epsilon}
\newcommand{\re}{r_{\mathcal{E}}}

\def\cM{{\cal M}}
\def\cV{{\cal V}}

\def\e{{\rm e}}

\newcommand{\zt}{\dot{z}}

\newcommand{\nul}{\, ^0  }
\newcommand{\un}{\, ^1  }
\newcommand{\de}{{\, ^2}  }
\newcommand{\pp}{\, ... \,}

\newcommand{\sgn}{{\rm sgn\smhsp}}
\newcommand{\nn}{\nonumber}

\newcommand{\br}{\mathrm{br}}
\newcommand{\p}{\mathrm{p}}

\newcommand{\ah}{ \mathrm{a}}
\newcommand{\bh}{ \mathrm{b}}

\newcommand{\smph}{\vphantom{ d^0_0}}
\newcommand{\vp}{\vphantom{\frac{a}{a}}}
\newcommand{\vph}{\vphantom{\frac{d}{d}}}
\newcommand{\Hh}{h}

\renewcommand{\geq}{\geqslant}
\newcommand{\kn}{{\kern 1pt}}
\newcommand{\akn}{{\kern -1pt}}

 \numberwithin{equation}{section}

\newcommand*{\hm}[1]{#1\nobreak\discretionary{}{\hbox{$\mathsurround=0pt #1$}}{}}

\begin{document}

\title{Piercing of domain walls: new mechanism of gravitational radiation}
\author{Dmitri Gal'tsov$^{a,b}$,
 Elena Melkumova$^{a}$  and
Pavel Spirin$^{a}$\footnote{E-mails: galtsov@phys.msu.ru, elenamelk@mail.ru,  pspirin@physics.uoc.gr}}

\affiliation{\parbox{12cm}{$^a$\kn{}Faculty of Physics, Moscow State University, 119991, Moscow, Russian Federation\newline
 $^b$\kn{}Kazan Federal University, 420008, Kazan, Russian Federation}}



\begin{abstract}
Domain wall  (DW) moving in media undergoes the friction force due to particle scattering. However certain particles are not scattered, but perforate the wall. As a result, the wall gets excited in the  form  of the branon wave, while the particle experiences an acceleration jump. This gives rise to generation of gravitational waves which we call ``piercing gravitational radiation'' (PGR). Though this effect is of higher order in the gravitational constant than the quadrupole radiation from the collapsing DWs, its amplitude is enhanced in the case of  relativistic particles or photons because of absence of the velocity factor which is present in the quadrupole formula. We derive the spectral-angular distribution of PGR within the simplified model of the weakly gravitating particle-wall system
in  Minkowski space-time of arbitrary dimensions.  Within this model the radiation amplitude is obtained analytically. The spectral-angular distribution of PGR in such an approach suffers from infrared and ultraviolet divergences as well as from collinear divergence  in the case of a massless perforating particle.  Different cut-off schemes  appropriate in various dimensions are discussed.  Our results are applicable both to cosmological DWs and to  the braneworld models.
\end{abstract}
\maketitle

\tableofcontents

\section{Introduction}

Since their prediction by Zeldovich, Kobzarev and Okun \cite{Zeldovich} and Kibble \cite{Kibble},
cosmological DWs  remain in the focus of theoretical study for more than forty years.  DWs are formed during the phase transitions in the early universe once the discrete symmetry of the underlying gauge theory is spontaneously broken. After creation, their average  number  per a Hubble radius remains constant for some time, so, if they were stable, their energy density potentially could dominate and overclose the Universe \cite{linear,linear1,Vilsh}.
To avoid this, DWs either must be unstable,  what happens if the discrete symmetry was only approximate, or disappear via some other mechanism. The basic viable field model of cosmological DWs is that of real scalar field with the biased potential  \cite{ki10}. Collapsing unstable DWs generate  gravitational waves, whose spectrum, sensible to particular underlying models,   can   be an important source of information about the early universe. In view of the forthcoming experimental studies of relict gravitational waves, this subject attracted  attention   recently
\cite{ki10,Kawasaki:2011vv,Hiramatsu:2013qaa,Kitajima:2015nla,Kadota:2015dza,Krajewski:2016vbr,Nakayama:2016gxi},
for a review see \cite{Saikawa:2017hiv}. Particular models considered include the hybrid inflation \cite{Kawasaki:2011vv}, the standard model extended to the very early universe \cite{Krajewski:2016vbr}, the Higgs model \cite{Kitajima:2015nla}, the next-to-minimal supersymmetric standard model \cite{Kadota:2015dza} and some other. The results are based on numerical simulations
of creation and annihilation of DW with account for gravitational radiation computed via the correlation functions of the energy-momentum tensors. These are in good agreement with the simple estimates based on the quadrupole formula for gravitational radiation of non-relativistic systems \cite{Nakayama:2016gxi}.

Collapsing DWs emit gravitational radiation due to their intrinsic dynamics, so the above mentioned numerical simulations seem to give major contribution to graviton production from the motion of the unstable DWs. These calculations, however, do not take into account  gravitational interaction of DW with  surrounding matter which may lead to additional generation of gravitational waves, maybe subdominant, but with distinct  spectral properties. This may be especially important if one assumes that the spontaneously broken discrete symmetry was exact, so no intrinsic instability of DWs is supposed.
Here we would like to discuss one such a mechanism which may be considered as generalization of the gravitational bremsstrahlung in particle collisions. We will be interested in perforating collisions of particles with extended objects. For more generality, we consider this process in any dimensions keeping in mind also the braneworld scenarios of the Randall-Sundrum II type \cite{RS1,RS2,Tanahashi:2011xx,BBHS3}. There the DW corresponds to factoring of the AdS5 geometry which leads to a particularly simple analytical description.

Recall that punctures of DWs, producing holes inside them, may serve as an alternative mechanism of DW destruction  \cite{ChamEard}. In the models admitting both the DWs and the cosmic strings, the hole of finite size in the wall must be surrounded by the string. Formation of such a hole changes the overall energy balance forcing the hole either to shrink due to tension of the string, or to expand due to DW tension, eventually eating the wall. More precisely, as was shown in  \cite{ChamEard}, at least {\em four}  holes are needed for the second option to be realized. The holes in the DWs could be created by bulk black holes perforating them \cite{Stojkovic:2005zh}, so physics of perforation is worth to be explored in detail. The field-theoretical treatment of the collision of black holes with DWs was developed in \cite{Flachi:2007ev}. These effects can be regarded as topological phase transitions. Other aspects of such transitions in the composite brane\,--\,black hole systems were studied in
\cite{BBHS, Frolov:2003mc, Czinner:2009rx, Czinner:2010hr, BBHS4}. In the context of the braneworld models, the interaction of black holes with DWs attracted much attention in connection with   the conjectured creation of  black holes  in particle collisions on the brane and their possible escape into the bulk \cite{escape,escape1,escape2}.  (In this framework we prefer to use the term ``brane'', though we always deal with branes of co-dimension one, that is with DWs.)

The problem considered here, though is related to the above issues, is, however, somewhat different.  We discuss perforation of DWs by elementary particles of the surrounding plasma (or bulk particles in the braneworld case) which classically have zero size, so our effects rather are dynamical than topological.  We use the framework first suggested in \cite{GaMeZ} and further developed in \cite{GaMeS1,GaMeS2,GaMeS3}, which essentially consists of perturbative description of gravitational interaction of the Nambu-Goto DW and the classical point particle in Minkowski space. This approach opens a way to describe analytically the branon excitation of the wall revealing creation of the free branon wave. This phenomenon was first found within the RS II five-dimensional model \cite{GaMeZ} and then generalized to arbitrary dimensions \cite{GaMeS1} showing that the branon effects are dimension-dependent  in view of the different causal structure of propagators of massless fields  in even and odd space-time dimensions \cite{GS}.  The particle-DW interaction has the growing potential, so there are no free asymptotic states. Still it turned out possible to introduce the dressed momenta for both objects which are instantaneously conserved during the collision \cite{GaMeS2}. We also calculated emission of scalar waves under perforation assuming that the particle (but not the wall) to interact  with a scalar field apart of gravity \cite{GaMeS3}.

Here we  calculate genuine gravitational radiation from the same system suggesting this as  new and mechanism of radiation: ``piercing gravitational radiation'' (PGR),  with some novel conceptual  and technical features. First, the system has no wave zone. This prompted us to revisit derivation of the spectral-angular distribution of radiated energy not referring to wave zone. Second, the effective radiating current contains a light-like part (due to the free branon) with associated problems in constructing the retarded solutions of the D'Alembert equation with light-like sources \cite{Azzurli:2012tp,Azzurli:2014lha,Tolish:2014bka}. Furthermore, radiation exhibits peculiar infrared and collinear   \cite{Kinoshita:1962ur,Lee:1964is,Nauenberg:1965uka,Weinberg:1965nx}  divergences,  typical for radiation from massless charges in gauge theories (for more detailed discussion see \cite{SR}).
Recently, these matters  were extensively studied in connection with
the memory effect
\cite{Tolish:2014bka,Tolish:2014oda,Bieri:2013ada,Winicour:2014ska,Bieri:2015yia}
and the Bondi-Metzner-Sachs asymptotic symmetries
\cite{Strominger:2014pwa}. Though we do not discuss these very interesting subjects here, we feel that our radiation problem provides a novel interesting setting
for these studies too.

Physically, PGR may be relevant for cosmic DWs, providing additional spectral components to the standard spectrum of gravitons from collapsing walls \cite{Saikawa:2017hiv}. Detailed study of such applications, however, remains beyond the scope of the present paper which is mostly restricted to theoretical aspects of the problem.

\section{The setup}

We consider  the gravitating system  of an infinite Nambu-Goto DW of plain topology and a point particle. Omitting the self-gravity of each object, we treat the full metric generated by them via Einstein equations and the motion of both objects in this metric self-consistently in the framework of the perturbation theory on the Minkowski background in terms of the  coupling constant $\vak$ (\mbox{$\vak^2=16\pi G_D$)}, where $G_D$ is the $D$-dimensional Newton constant (we use the units $c=1$). When gravity is switched off, the following geometry of the collision is assumed: the plane infinite Nambu-Goto brane sits at rest in  $D$-dimensional Minkowski space-time, so that its world-volume is orthogonal to $z$-axis. A point particle of mass $m$ is moving along $z$-axis with some initial velocity such that it reaches the wall and perforates it. The $D-$dimensional cartesian coordinates are therefore split as
\mbox{$x^M=(x^{\mu};z)$}, \mbox{$x^{\mu} \in M_{1,d+2}$}, the metric signature
is  \mbox{$(+,-,...,-)$}.
Our conventions for the Riemann
and Ricci  tensors  are: $R^B{}_{NRS}\equiv \Gamma^B_{NS , R} - \Gamma^B_{NR , S}
+ \Gamma^A_{NS} \Gamma^B_{AR} - \Gamma^A_{NR} \Gamma^B_{AS}$ and $R_{MN}\equiv
\delta^B_A\, R^A{}_{MBN}$.
\subsection{The model}
We   keep notation  introduced in the previous papers \cite{GaMeS1,GaMeS2,GaMeS3}.
The  $(D-2)$-dimensional DW propagating in the $D$-dimensional
space-time $\cM_D $  with the metric $g_{MN}$, has the world-volume
$\cV_{D-1}$  parametrized by arbitrary coordinates  $\sigma^\mu$,  $\mu=0,1,2,\,\pp \, , D-2$ and defined by the embedding equations
 $   x^{M}=X^{M}(\sigma^\mu)$,\; $M\hm{=}0,1,2,... ,D-1 $\,.
 The point mass
    propagates normally to the DW  along the worldline $x^M=z^M(\tau)$, affinely  parametrized and described by the Polyakov action using the einbein $e(\tau)$.
To total action governing the system reads
\begin{align}\label{Bac}
S = -\fr{\mu}{2}\int\left[\vp
  X_\mu^M X_\nu^N g_{MN}\gamma^{\mu\nu}-(D-3)\right]\sqrt{-\gamma}\,d^{D-1}
  \sigma  - \frac{1}{2} \int \! \left(e\; g_{MN}\dot{z}^M
\dot{z}^N+\frac{m^2}{e}\right) d\tau -\frac{1}{\vak^2}\!\int\! R_D
\,\sqrt{-g}\; d^D x \,,
 \end{align}
 Here $\mu$ denotes the brane tension, $X_\mu^M\hm{=}\pa
X^M/\pa\hsp\sigma^\mu$ are the tangent vectors on the DW world-volume and
$\gamma^{\kn\mu\nu}$ is the inverse metric on it,
$\gamma={\rm det} \gamma_{\mu\nu}$.

Variation of (\ref{Bac}) with respect to $X^M$ gives the brane
equation of motion in the covariant form
 \begin{align} \label{em}
\partial_\mu\left(  X_{\nu}^N
g_{MN}\gamma^{\kn\mu\nu}\sqrt{-\gamma}\right)=\frac{1}{2}\,
g_{NP,M}X^N_{\mu} X^P_{\nu}\gamma^{\kn\mu\nu}\sqrt{-\gamma}\,,
 \end{align}
while variation with respect to $\gamma^{\kn\mu\nu}$ gives the
constraint equation
 \begin{align} 
 \lbr X_\mu^M X_\nu^N - \fr12\,
\gamma_{\mu\nu}\gamma^{\la\tau} X_\la^M X_\tau^N \rbr g_{MN}
+\fr{D-3}{2}\,\gamma_{\mu\nu}=0\,, \nn
 \end{align}
 whose
solution defines $\gamma_{\mu\nu}$ as the induced metric on
$\cV_{D-1}$:
 \begin{align} \label{ceq} \gamma_{\mu\nu}=X_\mu^M  X_\nu^N g_{MN}{\big
 |}_{x=X}\,.
 \end{align}

Varying $S$ with respect to $e(\tau)$ and $z^M(\tau)$ one obtains
the equations
\begin{align}\label{consp}
e^2  g_{MN} \dot{z}^M \dot{z}^N=m^2
\end{align}
and
\begin{align}\label{eomp}
\fr{d}{d\tau}\lbr e \dot{z}^N g_{MN} \rbr=\fr{e}2 \;
g_{NP,M}\smhsp \dot{z}^N \dot{z}^P ,
\end{align}
respectively, while variation over $g_{MN}$ leads to the Einstein
equations
\begin{align}\label{Eeq}
G^{MN}=\frac{1}{2}\,\vak^2 \left[\vp T^{MN}+ \bar T^{MN}
\right]\,,
\end{align}
 with
\begin{align}\label{EMT}
T^{MN}=\mu\int X^M_\mu X^N_\nu
\gamma^{\mu\nu}\;\frac{\delta^D\left(x-X (\si)\smph
\right)}{\sqrt{-g}}\;\sqrt{-\gamma}\;d^{D-1}\si\, , \qquad
\bar T^{MN}= e \int \fr{\zt^M \zt^N \delta^D\!\left(x-z(\tau)\smph
\right)}{\sqrt{-g}}\,d\tau \,
 \end{align}
for the energy-momentum tensor of the brane and a particle,
respectively. Eventually in our expansions we will use  Einstein equations with lower indices.

Though the infinitely thin DW is still compatible with the full non-linear gravity,  the point-like particle is not. But it sensible to consider our system in the context
of the perturbative gravity on  Minkowski  background. In
this approach one presents the metric as
\begin{align}
g_{MN}=\eta_{MN}+ \vak \Hh_{MN}\,,\nn
\end{align}
and expands all quantities in powers of $\vak$, making use of
$\eta_{MN}$ to raise and to lower the indices.
It is convenient to define the quantity
\begin{align}
 {\psi}_{MN}\equiv \Hh_{MN}-\frac{\Hh}{2}  \, \eta_{MN}\,,
\qquad \Hh\equiv \eta^{MN}\,\Hh_{MN}\,, \nn
\end{align}
and to choose
 the flat-space harmonic gauge (in all orders in $\vak$)
\begin{align}
\label{hagef} \pa_N \psi^{MN}=0\,.
\end{align}
\subsection{Iteration scheme}
 Like in the particle scattering problem, we use the simultaneous expansions of the particle world-line $z^M(\tau)$, the DW embedding functions $X^M(\sigma^\mu)$ and the metric deviation  $\Hh_{MN}$ in $\vak$. It is understood that in  zero order
the particle moves freely along the line orthogonal to the wall and pierces it. The subtle point of this setting is that actually  the gravitational potential of the Nambu-Goto plane wall is growing in space, so the system has no free asymptotic states at all. Consequently one can not define the momentum of the particle in the initial state. This gives rise to complications in establishing the momentum balance equation. However, as was discussed in our previous paper \cite{GaMeS2}, there is a way out in constructing the ``dressed'' momenta, so we will not enter into this here. We still can use the formal expansions of the unknown functions in the gravitational constant  keeping in mind that we are actually considering a vicinity of the wall whose size is small with respect to curvature radius of an exact DW metric. Gravitational field of DW is repulsive, so to be able to pierce the DW, the particle must have enough energy in order not to be reflected. We will see in what follows that for any particle velocity there is some domain from which the particles reach DW. We will call this velocity dependent domain   {\em the piercing layer}, its size will be defined in the section 2C.  So actually our iterative scheme applies only to the piercing layer.

With this in mind, we proceed now  with formal expansions
\begin{align}\label{iterations}
\Phi=\nul\Phi+\un\Phi+{^2}\Phi+\ldots\,,
\end{align}
 where
$\Phi$ denotes collectively the set of variables: $z^M(\tau)$, $e(\tau)$,
$X^{\mu}(\sigma)$ and  $\Hh_{MN}(x)$. The left superscript
labels the order in terms $\vak$.

\vspace{0.3cm}

{\it The zeroth order} is trivial. It describes a free
plane unperturbed brane and a particle moving with constant
velocity ($u^M=\gamma(1, 0, ..., 0, v)$) perpendicular to the brane where
$\gamma=1/\sqrt{1-v^2}$:
\begin{align}
\nul z^\mu(\tau)= u^\mu\tau \,   , \nn
\end{align}
in the absence of the gravitational field
$h_{MN}=0\,.$
 The Lagrange multiplier  $e$  is chosen to be
equal to the corresponding particle mass $\nul e =m\, ,
$
 so that the trajectories are parametrized by
proper time  and the velocity satisfies the normalization
condition $\eta_{MN}u^M u^N\equiv u^2= 1$.

 In the zeroth order in $\vak$ the brane is
assumed to be unexcited
 \begin{align}
\nul \!X^M =\Si^M_\mu \si^\mu\,, \nn
 \end{align}
 where $\Si^M_\mu$  are $(D-1)$ constant bulk Minkowski vectors which can be
 normalized as
 \begin{align}
\Si^M_\mu\Si^N_\nu\eta_{MN}=\eta_{\mu\nu}\,, \nn
 \end{align}
so the corresponding induced metric on $\cV_{D-1}$ is flat.
Obviously, this is a solution to the Eq.~(\ref{em}) for $\vak=0$,
and the corresponding induced metric is four-dimensional Minkowski
metric $\gamma_{\mu\nu}=\eta_{\mu\nu}$. Thus it is convenient with no loss of generality to
fix
$$\Si^M_\mu=\delta^M_\mu \nn
\,.$$
In other words, we choose  the Lorentz frame
where the unperturbed brane is at rest.

 \vspace{0.3cm}

{\it The first order correction} is obtained next. The zeroth
order straight particle trajectory and the flat brane are sources
of the first order gravitational field $\un h_{MN}^{\p} \equiv
\bar{h}_{MN}$ of the particle and $\un h^{\br}_{MN}\equiv h_{MN}$ of the
brane, respectively:
\begin{align}
&\un h_{MN}=h_{MN}+\bar {h}_{MN}.\nn
\end{align}

In turn, $h_{MN}$ causes the first order deviation of the
trajectory $\un z^M$, while $\bar{h}_{MN}$ produces the first order
deviation $\un \! X^{\mu}(\sigma)$ of the embedding functions.
In the process, the first correction $\un e$ of the einbein fields
is also obtained. Explicitly, from the zeroth order trajectories
one obtains the zeroth order particle energy-momentum tensor
\begin{align}
\nul \bar{T}^{MN}=  m\int u^M u^N \delta^D\nhsp
\left(x-\nul z(\tau)\right)  d\tau\,, \nn
\end{align}
which in this order has only $t,z-$ components, and from the first
order Einstein equations for particle field, given  by
\begin{align}\label{hanuleq}
  \Box\, \bar {h}^{MN}=- \vak \left(\nul \bar{T}^{MN}-\frac{1}{D-2}  \nul \bar{T} \eta^{MN}  \right),
\end{align}
the first order correction $\un h^{MN}$ to the metric is obtained.
 In what follows, the stress-tensor and the gravitational field of
the particle will be denoted by bar.

 Passing to the $D$-dimensional Fourier-transforms
\begin{align}
   \bar h_{MN}(x)=\frac1{(2\pi)^D}\int \e^{-iqx}\bar h_{MN}(q)
\,d^Dq\,, \qquad\quad   \bar T^{MN}(x)=\frac1{(2\pi)^D}\int \e^{-iqx}
\bar T^{MN}(q)\,d^Dq\,, \nn
\end{align}
we obtain the retarded solution  in the momentum representation
 \begin{align}\label{ge_mom}
   \bar{h}_{MN}(q) =\frac{2\pi \vak m \, \delta(qu)}{q^2+i \varepsilon q^0} \left(u_M
u_N-\fr{1}{D-2}\,\eta_{MN}\right).
  \end{align}
In the coordinate representation we find (for $D\geq 4$):
\begin{align} \label{hpart}
 \bar{h}_{MN}(x)=-\fr{\vak
\,m\Gamma\!\left(\frac{D-3}{2}\right)}{4\pi^{\fr{D-1}{2}}}
 \left(u_M
u_N-\fr{1}{D-2}\,\eta_{MN}\right)\fr{1}{[\gamma^2(z-v
t)^2+r^2]^{\fr{D-3}{2}}}\,,
 \end{align}
where $r= \sqrt{\delta_{ij} \sigma^i \sigma^j}$ is the radial
distance on the wall from the perforation point. This is just the
Lorentz-contracted $D$-dimensional Newton  field of the  uniformly
moving particle.

The zeroth-order expression for the brane energy-momentum tensor
$T^{MN}$ reads:
 \begin{align}
\nul T^{MN}=\mu\int \Si^M_\mu\Si^N_\nu
\eta^{\mu\nu}\,\delta^{D}\!\left(x-\nul{\! X} (\si)\right)\,
d^{D-1}\si \,, \nn
 \end{align}
 so the Einstein equation for the first-order brane field is
 given by:
 \begin{align}
  \Box\, h^{MN}=- \vak \left(\nul
T^{MN}-\frac{1}{D-2} \nul T \eta^{MN}\right), \qquad \nul T \equiv
\nul T^{MN} \eta_{MN}\,. \nn
 \end{align}
The Fourier-space solution is given by
 \begin{align} \label{brgr00}
h_{MN}(q)= \frac{(2\pi)^{D-1}{\vak \mu}}{q^2} \left(\Xi_{MN}
-\frac{D-1}{D-2}\,\eta_{MN}\right)\delta^{D-1}(q^{\mu})\,,
 \end{align}
where $ \Xi_{MN} \equiv \Si_M^\mu\Si_{N}^{\nu} \eta_{\mu\nu}$,
while in the coordinate space it reads
 \begin{align} \label{brgr}
h_{MN}=\frac{\vak \mu}2\left(\Xi_{MN}
-\frac{D-1}{D-2}\,\eta_{MN}\right)|z|=\frac{\vak \mu
|z|}{2(D-2)}\;{\rm diag}\hsp(-1,1,\pp,1,{D-1})\,.
 \end{align}
\section{First order dynamics}
\subsection{Particle's motion}
Using $\un h^{MN}$ and the zeroth order solution in equations
(\ref{consp}) and (\ref{eomp}) one obtains for $\un e$ and $\un
z^M$ the equations\kn\footnote{Our gauge condition is $g_{MN} \dot
z^M \dot z^N=1$. To this order it reduces to $^1e=0$.}
\begin{align}\label{e1eq}
\un e=-\frac{ m}{2} \left( \vak
h_{MN}u^Mu^N+2\,\eta_{MN}u^M \un
 \dot{z}^N \vp \right)
\end{align}
and
\begin{align}\label{cucu0}
\fr{d}{d\tau}\lbr \un e u_M +m \un \dot{z}_M\rbr=-\vak m
 \left(h_{PM,Q}-\fr12\, h_{PQ,M}  \right) \;u^P u^Q\,,
\end{align}
which upon elimination of $\un e$ give
\begin{align}\label{cucu0_a}
     \un\ddot{z}_M =-\vak
 \left(h_{PM,Q}-\frac{1}{2}\, h_{PQ,M}  \right)u^P u^Q\,,
\end{align}
 or, in the components,
 \begin{align} \label{acce1}
 \un \ddot{z}^{\,0} =2kv\,
\gamma^2 \;{\rm sgn }(\tau)\,,\qquad \quad \delta\ddot z \equiv
 \un\ddot{z}^{D-1}=k\, (D\gamma^2 v^2+1)\;{\rm sgn } (\tau) \,,
 \end{align}
where a parameter is introduced
\begin{align}
k=
 \frac{\mu \varkappa^2}{4\left(D-2\right)}\nn
\end{align}
 playing the role of the inverse bulk curvature radius in the full non-linear treatment.
According to (\ref{acce1}) the gravity force between the DW and the particle is repulsive. (Recall that this ``antigravity'' is caused by the dominance of the DW tension (negative pressure) in the energy-momentum tensor of the wall.)

Integrating  (\ref{acce1}) twice with initial conditions $ \delta
z^M(0)=0, \;\delta\zt^M(0)=0$, one has
 \begin{align} \label{acce2}
 \un z^0   =k v \tau^2\, \gamma^2
\;{\rm sgn }(\tau)\,,\qquad \quad{ \delta z }=\frac{1}{2}\, k\tau^2
\left(D\gamma^2 v^2+1\right)\,{\rm sgn }( \tau)\,.
 \end{align}
Substituting  (\ref{acce2}) into (\ref{e1eq}) one can check that
the gauge condition $\un e=0$  is satisfied.
\subsection{Piercing layer}
We now  discuss physical restrictions  of  our iterative scheme in more detail.
Suppose that  a point  particle moving in the linearized gravitational field of the brane passes through the point \mbox{$z_l>0$} with the  velocity $dz/dt\hm{=}v\hm{<}0$.  According to Eq. (\ref{acce1}) it has a (proper) acceleration $\ddot{z} \hm{=}a\hm{=}k\, (D\gamma^2 v^2+1)$. This particle reaches the brane at the proper time moment
\begin{align}
\tau_0=\frac{|v|\gamma}{k(D\gamma^2 v^2+1)}\left(1-\sqrt{1-\frac{2kz_l(D\gamma^2 v^2+1)}{v^2\gamma^2}}\right).
 \end{align}
With account for the reflection symmetry, it is clear that if \mbox{$|z|>z_l$}, where
\begin{align}\lb{vdl}
z_l(v)=\frac{v^2\gamma^2}{2k(D\gamma^2 v^2+1)}\,,
\end{align}
 the particle will be reflected, while if  \mbox{$|z|<z_l$} it reaches the brane and passes through it. Thus the Eq.~(\ref{vdl}) defines the boundary of the velocity-dependent layer of particles which pierce the brane. Or, conversely, for given $z$, only those particles which have the velocity \mbox{$v>v_l(z)$}, where $v_l(z)$ is the inverse function to $z_l(v)$, namely,
\begin{align}\lb{zdl}
v_l(z)=\frac1{\gamma}\sqrt{\frac{2kz}{1-2kzD}}\,,
\end{align}
will reach DW and pierce it.

We will see in the next subsection that the parameter $k$  defines the curvature of the DW gravitational field. The linearized metric of the wall is correct if $kz\ll 1$ \cite{GaMeS1}.
In the non-relativistic case one has \mbox{$kz_l=v^2/2$}, so our approximation is consistent and the piercing layer is small in the units $k^{-1}$. In the ultrarelativistic case
$kz_l=(2D)^{-1}$, so the layer size is of the order of the inverse curvature. In this case the condition of validity of the linearized approximation for the brane metric   is only marginally satisfied.

Formally, the unperturbed motion is free in our scheme,  but one has to keep in mind that the validity of this description is restricted to  particles in the piercing layer only. Since this requirement is imposed a posteriori, one can expect that formal application of such an approach will face certain problems. Indeed, as we will see, the spectrum of gravitational waves  will require cut-offs to get finite results.

\subsection{The RS2 setup}

The piercing layer can be  described using the full non-linear treatment as follows. A convenient setup is the one-brane Randall-Sundrum model (RS II) \cite{RS1} adapted to arbitrary dimensions. Note that geodesic motion in the RS II setup was earlier considered with different motivation in \cite{geod1, geod2, geod3, Maziashvili:2005cd, Maziashvili:2005zp, Friedland:2008zz}\footnote{Actually the bulk in the RS II model is anti-deSitter, and the negative cosmological constant  is present, but for our local considerations restricted by the vicinity of the brane this is irrelevant}. The metric of the RS II model reads:
 \begin{align}
\mathrm{d}s^2=\e^{-2k|z|} \mathrm{d}s_\mathcal{M}^2-\mathrm{d}z^2\,, \nn
 \end{align}
where  $\mathrm{d}s_\mathcal{M}^2$ is the flat metric on the brane.
Using the non-vanishing Christoffel symbols
 \begin{align}
 \Gamma^0_{ij}=-\e^{-2k|z|} k \, \eta_{ij}  \, \sgn{z} \qquad \quad \Gamma^i_{jz}=-  k \,
 \delta^{i}_{j}  \, \sgn{z},
 \end{align}
in the geodesic equation $\ddot{x}^M + \Gamma^M_{LR} \dot{x}^L
\dot{x}^S=0$ one derives the following two equations
 \begin{align}
\ddot{t}=2 k \, \dot{t} \dot{z}\, \sgn{z}, \qquad\quad
\ddot{z}=\e^{-2k|z|} k \, \dot{t}^2 \,  \sgn{z}\,, \nn
 \end{align}
whose solution is
 \begin{align}
z(t) =\pm \frac{1}{2k}\ln\left(k^2 t^2 +\e^{2 k |z_0| }[1 \mp 2 k
v_0 t] \right),
 \end{align}
where $v_0=-(dz/dt)_{t=0}$.
The time $T$ needed for the particle  at an initial
distance $l$  to reach the brane is
 \begin{align}\label{Tgetbrane}
 T =\frac{1}{k}
\left[\e^{2 k l} v_0-\sqrt{e^{4 k l} v_0^2-\e^{2k l}+1}\right]\,,
 \end{align}
so we find  the following condition of piercing: the
lowest  initial value of the velocity $v_0$ has to be
 \begin{align}
v_0^{\min}=\e^{-2k l}\sqrt{\e^{2k l}-1}\,.\nn
 \end{align}
As a function of the product $kl$, the velocity $v_0^{\min}$ has the maximum
$1/2$ at $ k l =\ln 2/2$. Hence, if \mbox{$v_0>1/2$}, the particle
reaches the brane independently of the initial distance.
Conversely, for the fixed initial velocity $v_0$, the largest initial
distance $l^{\max}$ is
 \begin{align}
l^{\max}=\frac{1}{2k}\ln\left( \frac{1-\sqrt{1-4v_0^2}}{2v_0^2}
\right), \qquad v_0<\frac{1}{2}\,.
 \end{align}
Thus in our problem the non-relativistic limit implies either the Minkowski
limit of the metric, or an initial particle position on the brane.
In the non-relativistic limit one has
 \begin{align}
l^{\max}\simeq \frac{v_0 ^2}{2k}\,.\nn
 \end{align}
The particle velocity at the moment of piercing is given by
 \begin{align}
 v_{\rm br}\equiv \left.\frac{dz}{dt}\right|_{t=T}= -\sqrt{ e^{4 k
l} v_0^2-\e^{2k l}+1}\,,
 \end{align}
so the condition of  applicability of the perturbation theory is
 \begin{align}
 \frac{v_0-|v_{\rm br}|}{v_0} \ll 1\,.\nn
 \end{align}
Being translated to the possible values of $kl$, this gives
 \begin{align}
kl\ll 1,
 \end{align}
so
 \begin{align}
 \frac{v_0-|v_{\rm br}|}{v_0} \simeq kl\left( 2-\frac{1}{v_0^2}
 \right)\nn
 \end{align}
and $kl \ll 1$ is sufficient for the validity of our iterative scheme.
Hence, expanding $T$ (\ref{Tgetbrane}) in powers of $k$ one obtains
 \begin{align}
T \simeq \frac{l}{v_0}\bigl[1+\mathcal{O}\kn(kl)\bigr],
 \end{align}
what corresponds to the linear gravity. In order for this would give the limit
of the full non-linear theory, where particle is replaced by the black hole, one
has also to ensure that we deal with distances large compared with the gravitational radius:
 \begin{align}
l\gg r_g\,, \quad {\rm or} \quad m\ll \vak^2 \mu^2\,\nn
 \end{align}
\subsection{Deformation of  DW}\label{Branon_wave_equation}
Now we consider perturbations of  DW due to
gravitational interaction with the  particle. For this
we have to use the metric deviation due to the particle. In
accordance with our iterative scheme we
neglect particle's acceleration in the wall gravity when we
calculate its proper gravitational field, considering the
unperturbed particle trajectory.

Perturbations of the Nambu-Goto branes in the external gravitational
field were expensively studied in the past, see e.g.
\cite{Gu}. The derivation is
particularly simple in the Minkowski background . First, from Eq. (\ref{ceq}) we find the
perturbation of the induced metric
 \begin{align}
  \delta \gamma_{\mu\nu}=2\hsp  \Sigma^M_{(\mu}\hsp \delta\nhsp \smhsp X^N_{\nu)}\eta_{MN}+
 \vak \bar h_{MN}\Sigma^{M}_{\mu}\Sigma^{N}_{\nu} \,, \nn
 \end{align}
where brackets denote symmetrization over the indices with a factor
$1/2$. Linearizing the rest of the Eq. (\ref{em}), after some
rearrangements one obtains the following equation for the deformation
of the wall:
 \begin{align}  \label{pisk}
 {\Pi}_{MN}\;\Box_{D-1}\;\delta\nhsp \smhsp
X^N= {\Pi}_{MN}\;J^N\,, \qquad  {\Pi}^{MN} \equiv \eta^{MN}-
\Sigma^{M}_{\mu}\Sigma^{N}_{\nu} \eta^{\mu\nu}\,,
 \end{align}
 where $\Box_{D-1} \equiv \partial_{\mu} \partial^{\mu}$ and $ {\Pi}^{MN} $ is
projector onto the (one-dimensional) subspace orthogonal to
$\cV_{D-1}$. The source term   in  (\ref{pisk}) reads:
 \begin{align} \label{JN}
 J^N=   \vak \, \Sigma_P^\mu \,\Sigma_Q^\nu\,
\eta_{\mu\nu} \left(\frac{1}{2} \, \bar h^{PQ,N} -\bar
h^{NP,Q}\right)_{\!z=0}\! .
 \end{align}
Using the aligned coordinates on the brane
$\sigma^{\mu}=(t,\mathbf{r}), $ we will have
$\delta^{M}_{\mu}=\Sigma^{M}_{\mu}$, so  the projector $
{\Pi}^{MN}$ reduces the system (\ref{pisk}) to a single equation
for the $M=z$ component. Thus only the $z$-component   of $\delta\nhsp
\smhsp X^M$ and $J^M$ is physical. Generically, the transverse
deformations of branes can be viewed as Nambu-Goldstone bosons
(branons) which  result from spontaneous breaking of the
translational symmetry \cite{KuYo}. In the brane-world models these
are coupled to matter on the brane  via the induced
metric (for a recent discussion see \cite{Bu}). In our case
of co-dimension one there is only one such branon. The remaining
components of the perturbation $\delta\nhsp \smhsp X^M$ can be
removed by  transformation of the coordinates on the
world-volume, so $\delta\nhsp \smhsp X^\mu=0$ is nothing but the gauge
choice. Note that in this gauge the perturbation of the
induced metric $\delta \gamma_{\mu\nu}$ does not vanish, contrary to
the perturbation of the particle ein-bein $e$.

Denoting the physical component as $\Phi(\sigma^\mu) \equiv
\delta\nhsp \smhsp X^z$ we obtain the branon $(D-1)$-dimensional
wave equation:
 \begin{align} \label{NGEQ}
\eta^{\kn\mu \nu}\frac{\partial}{ \partial \sigma^{\mu}}\frac{\partial}{ \partial \sigma^{\nu}} \, \Phi(\si)=J(\si),
  \end{align}
  with  the source term $J \equiv J^z$.
Substituting (\ref{hpart}) into the eq.\,(\ref{JN}) we obtain the
source term for the branon:
 \begin{align} \label{jxb}
 J( \sigma )=-\vak \left[\frac{1}{2}\,  \eta_{{\mu\nu}}\bar{h}^{\hsp \mu\nu,z}-\bar{h}^{\hsp  z\hsp  0,0}\right]_{z=0} =-  \fr{\la vt}{[\gamma^2 v^2
t^2+r^2]^\fr{D-1}{2}}\,,\end{align}
 where
\begin{align}
\la=\fr{\vak^2 m\gamma^2\Gamma\left(\fr{D-1}{2}
\right)}{4\pi^{\fr{D-1}{2}}}\left( \gamma^2v^2 +\fr{1}{D-2}\right)
.\nn
 \end{align}

Construction of the retarded solution of the eq.\,(\ref{NGEQ}) was explained in \cite{GaMeZ,GaMeS1}, the result consists of two terms:
  \begin{align}\label{Phi11}
\Phi(t, \mathbf{r})=  - \Lambda \,\sgn\kn(t)\,I_{\ah} + 2\, \Lambda  \, \theta\kn(t)\,I_{\bh}\,,\qquad \Lambda \equiv
\frac{ \sqrt{\pi}
 \,\la}{2^{ \frac{D-2}{2}} \gamma^3\Gamma\left(\fr{D-1}{2}
 \right)}\,,
\end{align}
with
\begin{align}
  I_{\ah}(t,r) = \frac{1}{r^{\fr{D-4}{2}}} \int\limits_{0}^{\infty} \! dy
\,J_{\frac{D-4}{2}}(y r)\,{y}^{\frac{D-6}{2}} \, \e^{- y\gamma
v|t|}\,,    \qquad \quad
 I_{\bh}(t,r)= \frac{1}{r^{\fr{D-4}{2}}} \int\limits_{0}^{\infty} \!
dy\,J_{\frac{D-4}{2}}(y r)\,{y}^{\frac{D-6}{2}} \,\cos \, yt\,
,   \nn
 \end{align}
from which the first describes  the odd in time brane deformation caused by the Lorentz-contracted Newton field of the particle, while the second is the shock branon wave arising at the moment of  perforation and  then freely propagating outwards along the brane. For \mbox{$D=4$} these two integrals  diverge logarithmically, though, as it was shown in \cite{GaMeS1}, the corresponding regularized solutions exist. Here we will use the direct solution of the Eq.~(\ref{NGEQ}) in the momentum representation:
 \begin{align} \label{NGEQ_mom_f1}
 \Phi\kn(q)=-\frac{i\kn \pi \kn \vak^2 m }{\gamma}
  \frac{ q^z\, \delta(q^0-v\,q^z)}{q_{M}q^{M} (q_{\mu}q^{\mu}+2\kn i\kn \ep q^0)}  \left( \gamma^2 v^2+\frac{1}{D-2} \right).
  \end{align}
\section{Gravitational radiation formula revisited}
Traditionally, both electromagnetic and gravitational radiation is computed in terms of fluxes of the field momentum in the wave zone, which is well-defined only in the asymptotically flat space-time. Our space-time is not asymptotically flat, so  one
should revisit the derivation. In particular, the energy-momentum flux through the lateral surface of the world-tube turns out to be non-zero \cite{GaMeS3}. Meanwhile, one can transform the flux at infinity, when it is well defined, into the volume integral extended through the space-time. This allows one to express the radiation power as an integral over the graviton momentum from the square of the source stress-tensor in the momentum representation contracted with polarization tensors \cite{Weinberg}. Here we  present the derivation of essentially the same formula without reference to the wave zone.

\subsection{The second order}
 In the second order in $\vak$ one obtains the leading
contribution to gravitational radiation. Actually, the source of radiation consists of three ingredients. The first is due to the particle which has constant acceleration before and after piercing.  This   has certain analogy with the  Weinberg's
approach \cite{Weinberg} to compute gravitational radiation from the system of   particles colliding at a point: in that case one has the constant {\em momenta} before and after collision which instantaneously change on a finite amount. In our case it is the (proper) time {\em derivatives} of the momenta before and after   collision which  are constant and opposite, changing sign at the moment of perforation.
The second contribution comes from the deformation of the brane world-volume caused by varying gravitational field of the moving particle. Recall, that in our setting the brane is plane and non-excited once gravitational interaction is switched off. Finally, for consistency of calculations, the gravitational stresses have to be taken into account, these are described using Weinberg's  expansion of the Einstein tensor up to the second order in the gravitational constant \cite{Weinberg}. The gravitational stresses constitute the third component of the source.

The Einstein   equation expanded up to the second order together with the corresponding expansion of the metric leads to the following equation for the second-order (trace-reversed) metric deviation:
$^2\psi_{MN}$:
\begin{align}\label{psi2eq} \Box\, \de\psi_{MN}=-\vak\;\tau_{MN}\,,
\end{align}
with the source  containing three terms:
\begin{align}\label{source2}
\tau_{MN}= \un \bar {T}_{MN}+\un T_{MN}+S_{MN}\bigl( \un h
\bigr)\,.
\end{align}
The first is the particle term following  from  the Eq.~(\ref{EMT}):\kn\footnote{Symmetrization over two
indices is defined according to $A_{(MN)}\equiv
(A_{MN}+A_{NM})/2$.}
\begin{align} \label{T1MN} \un \bar {T}_{MN}(x)=  m \int   \left[ 2\un \dot z_{(M}
u_{N)} +\vak \lbr 2u^P h_{P(M}u_{N)} -\frac{h}{2}\, u_M
u_N\rbr - u_M u_N \un {z}^P\pa_P \right] \delta^D\!(x-\! \! \nul
z(\tau))\, d\tau\,.
\end{align}
The second term $\un T_{MN}$
represents the brane contribution. To compute it, one substitutes the
first-order metric deviation (\ref{hpart}) and the first-order brane
perturbations into the Eq.~(\ref{EMT}), keeping  the                                                                                                                                                                                                                                                                                                                                                                                                                                                                                                                                                                                                                                                                                                                                                                                                                                                                                                                                                                                                                                                                                                                                                                                                                                                                                                                                                                                                                                                                                                                                                                                                                                                                                                                                                                                                                                                                                                                                                                                                                                                                                                                                                                                                                                                                                                                                                                                                                                                                                                                                                                                                                                                                                                                                                                                                                                                                                                                                                                                                                                                                                                                                                                                                                                                                                                                                                                                                                                                                                                                                                                                                                                                                                                                                                                                                                         quantities of the desired order:
 \begin{align} \label{taumn}
 \un T_{MN}  =
\fr{\mu}{2} \int & \left[ 4\, \Sigma^{\mu}_{(M}\un{\!X}^{\vm{\mu}
\nu}_{N)} \eta_{\mu\nu}+4\,\bar{h}_{
\lambda\,(M}^{\vphantom{\lambda}}\Sigma^{\lambda}_{N)}   -2\,
\Sigma_M^\mu \Sigma_N^{\nu\vm{\mu}} \left( \bar{h}_{ \mu\nu}  + 2\,
\eta_{LR} \Sigma^R_{( \mu } \un{\!X}^L_{\nu)} \right) + \right.
\nn\\ & \left. \;\;+
 \Sigma_M^{\mu}\Sigma_{N}^{\nu} \eta_{ \mu\nu} \left(\bar {h}^{\lambda}_{\lambda}-
{\bar {h}}+2\un{\!X}^{L}_{\lambda}\Sigma^{\lambda}_L -2\un{\!X}^{L}
\partial_L \vp\right) \right]\delta^D\!(x^A- \Sigma^A_\al\sigma^\al)\, d^{D-1}\si \,,
 \end{align}
where ${\rule{0em}{0.8em}\Sigma}^{\al}_{\nhsp M} \equiv
\Sigma^{N}_{\nu} \eta^{\nu \al} \eta_{MN} $ (and the similarly for
$\un{\!X}_{\mu}^{M}$).

To construct the stress term one uses the expansion of the Einstein tensor in
powers of $h_{MN}$   in the harmonic gauge:
\begin{align}
\label{SMN} G_{MN}=-\fr{\vak}{2}\, \Box \psi_{MN}-\fr{\vak^2}{2}\,
{\sf{S}}_{MN}+ \mathcal{O}\kn(\vak^3),
\end{align}
where $\Box=\pa_M\pa^M$ is the flat D'Alembert operator, ${\sf
S}_{MN}$ is the $\mathcal{O}\hsp(h^2)$ part of
$G_{MN}$
given by
\begin{align}
\label{natag_0} {\sf{S}}_{MN} [\Hh] =& \,  \Hh_M^{P ,
Q}\left(\Hh_{NQ , P} - \Hh_{NP , Q}\vp \right)
+\Hh^{PQ}\left(\Hh_{MP ,NQ}+ \Hh_{NP , MQ}- \Hh_{PQ, MN}- \Hh_{MN
, PQ}\vp \right) -\\
  & -\frac{1}{2}\, \Hh^{PQ}_{\quad , M} \Hh_{PQ ,
N}-\frac{1}{2}\,\Hh_{MN}\Box \Hh +
\frac{1}{2}\,\eta_{MN}\!\left(2\Hh^{PQ}\Box \Hh_{PQ}-\Hh_{PQ , L}
\Hh^{PL , Q}+\frac{3}{2}\, \Hh_{PQ , L} \Hh^{PQ,L}\right). \nn
\end{align}
In the source term in Eq.~(\ref{psi2eq}) one has to use the squares of the first-order quantities $\un h_{MN}$ in ${\sf{S}}$.
Furthermore, ignoring self-interaction, one has to keep in ${\sf {S}}_{MN}(\un h)$ only the products of the first order metric deviations $\un h$ {\em due to the
particle and the brane}. In other words, in ${\sf{S}}_{MN}$ we retain
$h\bar {h}$-terms and not the terms $\bar {h}\bar {h}$ and $hh$
responsible for self-action.

Using the equations for the first-order fields it is straightforward to verify that
\begin{align}
\pa_N \tau^{MN}=0\,,\nn
\end{align}
which guarantees the validity of the gauge fixing condition
(\ref{hagef}) to this order.

Thus ${\sf{S}}_{MN}$ considered as a quadratic form in $\un h_{MN}$
constitutes the non-local (in terms of the flat space picture)
source of gravitational radiation. This non-locality is due to the
non-linearity of the underlying full Einstein theory and, as will be shown, it leads to an
important difference  in the radiation spectrum compared with the results of linear
theories like electromagnetism. More detailed discussions of this
point  can be found in \cite{Galtsov:1980ap} within the four-dimensional theory, and in arbitrary dimensions, but within  a
simpler  scalar model, in \cite{GKST-3}.

\subsection{New derivation}

One starts with the particle  equation of motion in an external gravitational field  (\ref{cucu0_a}) written in terms of the covectors:
\begin{align}\label{EOM0}
 \ddot{z}_{M}  =\frac{1}{2}\, g_{AB, M} \zt^A
 \zt^B\, .
\end{align}
The idea is to present the change of the particle momentum $\Delta \bar{P} _{M}$ (eventually we will consider the full time of motion) as the integral over the entire space-time. This is done passing to the energy-momentum tensor:
\begin{align}
\Delta \bar{P} _{M}=-  m \int \ddot{z}_{M} ds =-\frac{1}{2}\int
g_{AB , M} \bar{T}^{AB} \sqrt{-g} \, d^D x\,.\nn
\end{align}
Similarly we can present the change of the momentum of the wall, this quantity is assumed to be finite, while the momentum itself is infinite (for more detailed discussion see \cite{GaMeS3}). Using the Eq.~(\ref{em}) we obtain:
 \begin{align} \label{em2}
\Delta P_{M}&=- \mu \int \partial_\mu\bigl(  X_\nu^N
g_{MN}\gamma^{\mu\nu}\sqrt{-\gamma}\bigr)\,d^{D-1}\sigma\nn \\ & =-\frac{
\mu}{2}\int \, g_{AB,M} \kn \Si^{AB} \sqrt{-\gamma}\,d^{D-1}\sigma \nn \\ & =
-\frac{1}{2}\int g_{AB , M} \kn T^{AB} \sqrt{-g} \, d^D x,
 \end{align}
Denote the total matter energy-momentum tensor  ${\sf{T}}^{MN} \equiv \bar{T}^{MN}
+T^{MN}$.
Since ${\sf{T}}^{MN}$ is a symmetric conserved tensor, one has
\begin{align}
 {\sf{T}}_{M; \,A}^{A}=\frac{1}{\sqrt{-g}}\,\bigl({\sf{T}}_{M}^{A} \sqrt{-g}\bigr)_{,A} - \frac{1}{2} \, g_{AB , M}
 {\sf{T}}^{AB}=0\,.\nn
\end{align}
The change of the total momentum $\Delta {\sf P}_{M}$ will read:
\begin{align}
\Delta {\sf P}_{M} =- \int \bigl({\sf{T}}_{M}^{A} \sqrt{-g}\bigr)_{,A}  \, d^D x =- \int
\Bigl({\sf{T}}_{M , A}^{A} \sqrt{-g} + {\sf{T}}_{M}^{A}\sqrt{-g}_{\akn ,A} \Bigr) \,
d^D x\,,\nn
\end{align}
or, equivalently,
\begin{align}\label{RRF1ss}
\Delta {\sf P}_{M} = - \int \left[{\sf{T}}_{M , A}^{A}+\frac{1}{2}
 {\sf{T}}_{M}^{A} \eta^{NL} g_{NL ,A} \right]\sqrt{-g} \, d^D
x\, .
\end{align}

Now we proceed with the expansions (\ref{iterations}) in terms of $\vak$
for both the stress-energy tensor ${\sf{T}}_{M}^{A} $ and the metrics $g_{NL}$.
Substituting   (\ref{iterations}) for $\Hh$ into
(\ref{SMN}), one obtains the expansion of the Einstein tensor, from which we retain terms up to $\vak^2$:
\begin{align}\label{high}
G_{MN}=&  -   \frac{\vak}{2}\Box \un \psi_{MN} + \left[ - \frac{\vak
}{2}\Box  \de \psi_{MN} - \frac{\vak^{2} }{2}{\sf{S}}_{MN}(\un h )\right]\,.
\end{align}
The Einstein equations in the two lowest orders read:
\begin{align}\label{natag_0a}
 & \Box\un \psi_{M}^{N}=-\vak  \nul {\sf{T}}_{M}^{N} \\ \label{natag_0b} & \Box \de
 \psi_{M}^{N}=-\vak\,
 \left(\!\un {\sf{T}}_{M}^{N}+{\sf{S}}_{M}^{N}\right)\,.
\end{align}
The sum $\tau_{M}^{N} \equiv \un {\sf{T}}_{M}^{N}+{\sf{S}}_{M}^{N}$
has zero divergence in accordance with the gauge choice in each order of
 $\psi_{MN}$.
Recall again that in all products of the first order metric deviations  one has to keep only the crossed particle-brane terms.

Now we have to carefully analyze the relevant orders in (\ref{RRF1ss}). One can  drop the first    $\vak$-order terms  in $\tau_{M}^{N}$  keeping in mind that radiation  in the self-gravitating systems can arise only in the second  order,
 so the relevant order of the metric derivative is
$ \eta^{NL} g_{NL ,A} $  is $ \vak \de h_{A} $. The lowest
order  of $T_{M, A}^{A}$  is $\un T_{M,A}^{A}$ by virtue of  $\nul
{\sf{T}}_{M,A}^{A} =- \Box \un \psi_{M, A}^{A}=0 $ by the gauge fixing.
Finally,  the lowest order  of $T_{M }^{A}$  is $\un {\sf{T}}_{M }^{A}$
due to (\ref{natag_0a}) and the asymptotic behavior  of $\Box
\psi$, as described in details in \cite{GKST-4}.
Therefore:
\begin{itemize}
    \item instead of ${\sf{T}}_{M , A}^{A}$ in (\ref{RRF1ss}) we write  $2\, G_{M , A}^{A}/
    \vak^{2}$, then substitute   (\ref{high}),
    eliminate all $ \Box  \,(^{k+1}\psi_{M , A}^{A})=0$ and all terms
    with $\un h$ as multiplier, this operation will be denoted by $\simeq$, hence
\begin{align}
{\sf{T}}_{M , A}^{A}\simeq - { {S}}_{M , A}^{A}(\de h )\quad
\text{plus higher order terms}\nn
\end{align}
    \item the same procedure applied to ${\sf{T}}_{M}^{A} \eta^{NL} g_{NL ,A}$, gives
\begin{align}
\frac{1}{2}
 {\sf{T}}_{M}^{A} \eta^{NL} g_{NL ,A}\simeq  - \frac{1}{2}   \de h_{,A}
 \Box \de \psi_{M}^{A} \quad \text{plus
higher order terms.}\nn
\end{align}
\end{itemize}

Substituting this into (\ref{RRF1ss}) one gets
\begin{align}\label{RRF2}
\Delta P_{M} =   \int \left[{ {S}}_{M , A}^{A}(\de h)+\frac{1}{2}  \de  h_{,A}  \Box \de \psi_{M}^{A}
 \right]\sqrt{-g} \, d^D x
\end{align}
where ${\sf {S}}_{M}^{N}$ is mixed indices tensor  in the relevant order\kn\footnote{We use mixed tensors just to shorten the formulas, actually we compute
the  expansion of $G_{M N}$ with \emph{lower} indices.}:
\begin{align}
{{S}}_{M}^{N}=-\frac{1}{\vak^2}\left(2\,G_{M}^{(2)N}+\vak\, \Box
\psi_{M}^{N}\right).\nn
\end{align}
Eventually (also omitting all $\un h$-terms):
$${{S}}_{M}^{N}(\de  h, \de  h) =\eta^{NL}{ {S}}_{ML}(\de  h) - \de
h^{NL}  \Box \de \psi_{ML}.$$
Finally, since the expression in brackets in (\ref{RRF2}) become the square of the second order terms (our desired order), we put the metric determinant  $\sqrt{- g}=1$.

Now we calculate the divergences
$$\partial^{N}{ \sf {S}}_{MN}(\de  h) =\frac{1}{2}  \, \Box \de  \psi^{AB} \de h_{AB,M} -\frac{1}{2} \de  h_{MN}  \Box\de h^{,N}
+ \de  h^{AB} \Box  \de  h_{MA,B}$$ and
$$ \partial_{N} ( \de  h^{NL} \Box \de \psi_{ML})=\frac{1}{2} \de  h^{,L}\Box \de  \psi_{ML} + \de h^{NL} \Box
\de h_{ML,N}-\frac{1}{2}\de  h_{MN} \Box \de   h^{,N}$$ and
substitute them into (\ref{RRF2}),  arriving at
\begin{align}\label{RRF3}
\Delta P_{M} =  \frac{1}{2} \int \de  h_{AB,M} \Box \de \psi^{AB}
 \, d^D x
\end{align}
which is the same formula as usually derived using the wave zone considerations   (see, e.g. \cite{GKST-4}). Replacing here  $\Box \de \psi^{AB}$
by the source terms in the corresponding wave equation, one can interpret this integral as the work done by the radiation reaction force. So this approach is still valid in our problem, where the notion of the wave zone is problematic. We also address the reader to our previous paper \cite{GaMeS2} for discussion of the dressed conserved quantities in the case without free asymptotic states.

Finally, repeating the same steps as in \cite{GKST-4}, and taking $M=0 $
one ends up with the standard expression for the radiated energy in terms of the Fourier-transform of the effective source:
\begin{align}\label{epol}
E_{\rm rad} =\frac{\vak^2}{4(2 \pi)^{D-1}}\sum\limits_{\cp}
\int\limits_{0}^{\infty}\omega^{D-2} d\omega \!\!\!
\int\limits_{{S}^{D-2}} \!\!  d\Omega \,
    \left| \varepsilon^{SN}_{\cp}\tau_{ SN}(k)\right|^{2} \, ,
\end{align}
 where  $ \{\varepsilon ^{\cp} \}$
represents the set of $ D(D-3)/2$ polarization tensors, to be
constructed below. Note that this formula gives the total radiation energy, not  the radiation power.
\subsection{Polarization tensors}\label{polarz}
Here we construct the polarization states in the gauge convenient to further calculations. In $D$
dimensions there are $D(D-3)/2$ independent second-rank symmetric tensors
$\varepsilon^{\cp}$, satisfying the following conditions:
\begin{itemize}
    \item transversality:
 \begin{align}\label{bark}
k^{N} \varepsilon_{MN}^{\cp}=0\,, \qquad \bar{k}^{N}
\varepsilon_{MN}^{\cp}=0\, ,
  \end{align}
 where $k^M=(\omega, {\bf
k}),$ $\bar k^M=(\omega, -{\bf k})$;
    \item tracelessness:
 \begin{align}
\eta^{MN} \varepsilon_{MN}^{\cp}=0\, ;\nn
  \end{align}
    \item orthonormality:
 \begin{align}
  \varepsilon_{MN}^{\cp} \varepsilon^{MN\;\cp'}=\delta^{\cp\cp'}\,.\nn
  \end{align}
\end{itemize}
To proceed, we  first construct  $D-2$ space-like unit vectors $e_{a}^{M}$ orthogonal to $k$ and $\bar{k}$ and\kn\footnote{This also implies orthogonality
to the time direction vector $(k+\bar{k})/2\omega =(1,0, \pp, 0)$.} among themselves:
 \begin{align}\label{eparity}
\eta_{MN}e_{a}^{M} e_b^{N}=-\delta_{ab},\qquad  k_M e^{M}_
{a}=0,\qquad \bar k_M e^{M}_{a}=0\,.
 \end{align}
To specify them further, introduce the unit space-like vector $\underline{n}$  $(\underline{n}^2=-1$), associated with the unit sphere $S^{D-3}$ within the DW, and the angle $\psi$ between  ${\bf k}$  and the $z$-axis (the line of particle motion). Then the graviton wave-vector will be parametrized as
 \begin{align} \label{k}
   {\bf k}= \omega\,( \underline{n}\,\sin\psi, \cos\psi).\end{align}
    Next parametrize the sphere $S^{D-3}$ by $(D-4)$ polar angles
   $\theta_1,\theta_2,\, ... \, ,\theta_{D-3}$ running from 0 to $\pi$ and one azimuthal angle
   $\phi$ running from 0 to $2\pi$.
Then the desired orthogonal $D-2$ unit vectors  will consist of the following two
 \begin{align}
{e}^\psi = (0,\underline{n}\cos\psi , -\sin\psi), \qquad
{e}^\phi =(0,\pa_\phi \underline{n}, \; 0)\nn
\end{align}
and the $D-4$ vectors
\begin{align}
{e}^{i}= (0,\pa_{\theta_i} \underline{n}, \; {0})\, .\nn
\end{align}
They all satisfy the conditions (\ref{eparity}) and have no temporal
component by construction. From them only the vector $e^{\psi}$ has the bulk component, while the remaining vectors belong entirely to the wall and  satisfy the mutual orthogonality conditions:
$$\eta_{\mu\nu}e_{a}^{\mu} e_b^{\nu}=-\delta_{ab}\, , \qquad \delta_{ij}e_{a}^{i} e_b^{j}=\delta_{ab} \, , \qquad a,b=\phi,\theta_i\,.$$

Now, using the above unit vectors, define the following two sets of symmetric polarization tensors:
\begin{itemize}
    \item $(D-2)(D-3)/2$  tensors labeled by pairwise indices $rs$:
    \begin{align}\label{tip1}
\varepsilon_{MN}^{rs}= \fr{e_M^r e_N^s +e_M^s e_N^r}{\sqrt{2}}
\qquad r,s =\phi, i, \psi\,,
\end{align}
    \item
 $D-3$ tensors labeled by a single index $\alpha=i,\, z$
 \begin{align}\label{tip2}
 \varepsilon_{MN}^{\alpha} = \frac{1}{\sqrt{\alpha(\alpha+1)}}
 \biggl[   \alpha \,e_M^\alpha  e_N^\alpha-\sum_{\beta=1}^{\alpha-1}e_M^\beta e_N^\beta-e_M^\phi e_N^\phi
 \biggr]\,,
 \end{align}
where the last one can be rewritten as
 \begin{align}\label{tip3}
 \va^z_{MN}=\fr{(D-3)\,e_M^\psi e_N^\psi-\sum\limits_{j=1}^{D-4}e_M^{j}e_N^j-e_M^\phi e_N^\phi}
 {\sqrt{(D-3)(D-2)}}\,.
 \end{align}
It is the only one  possessing  the bulk indices via  the $e_M^\psi e_N^\psi$-term.
\end{itemize}
For future purposes, it is worth noting that the polarization tensors $\va^{i\psi}, \va^{\phi\psi}$ and $\va^{z}$, with $i =\phi, \theta_i$  contain bulk indices ($z$-directed) through the vector $e_M^\psi$, while the remaining  $\va^{ij}$,
and $\va^{i}$  do not.
\section{Radiation amplitudes}\label{Radamp}
According to Eq.~(\ref{epol}) we have to compute the polarization projections of the effective tensor current on the graviton mass-shell, $k^2=0$.  The total current $\tau^{\kn\mu\nu}$ (\ref{source2}) consists of three terms: the brane term $\un T_{MN}(k)$, comprising contribution of the reversible deformation due to the variable gravitational field of the moving particle and the shock branon wave arising at the moment of piercing;  the particle term $\un \bar{T}_{MN}(k)$, encompassing corrections to the free motion due  to the gravitational field of the brane;  and   ${S}_{MN}(k)$, the bilinear product of $h_{MN}$ and $\bar{h}_{MN}$, which is the  stress-tensor of the gravitational field. It is worth noting that though the sum of the three terms is uniquely defined up to longitudinal  $k^M$ terms vanishing under polarization projection, each separate contribution  is gauge dependent, so the following calculation of separate  terms  is associated with specific gauge choices for the brane and the particle which were described above and will be commented again later.
\subsection{The brane amplitude}
The first-order  brane stress-tensor in the coordinate
representation is given by (\ref{taumn}). Specifying the variations of the world-volume embedding functions as $\un{\!X}^{N} = \Phi(\sigma)\,\delta^{N}_z $  and passing to the Fourier-transform we obtain:
 \begin{align} \label{taumn4}
 \un T_{MN}(k)  =
\fr{\mu}{4\pi} \!\int \!& \left[ -4\,i k_{\la}\Phi \kn(q)\,
  \Sigma^{\lambda}_{(M} \eta_{N)z}^{\vm{\lambda}} +4\,  \Sigma^{\lambda}_{(M}  \bar{h}_{N)\lambda}^{\vphantom{\lambda}}(q)  -2\, \Sigma_M^\mu
\Sigma_N^{\nu\vm{\mu}}  \bar{h}_{ \mu\nu}(q)   + \right. \nn \\ & \;\;
\left. +\,
 \Xi_{MN}\! \left(\bar{h}_{zz}(q) -2\,i\, k^z \Phi\kn(q)
\vp\right) \right]\delta^{D-1}(k^{\mu}-q^{\mu})\, d^{D}q\,,
 \end{align}
where $ \Phi(q^{M})$ is defined by the Eqs. (\ref{JN}), (\ref{NGEQ}) and
(\ref{jxb}) and reads
 \begin{align} \label{NGEQ_mom_f}
 \Phi\hsp(q^{M})=-\frac{i\, \pi \, \vak^2 m }{\gamma}
  \frac{ q^z\, \delta(q^0-v\,q^z)}{q_{M}q^{M} (q_{\mu}q^{\mu}+2i\ep q^0)}  \left( \gamma^2 v^2+\frac{1}{D-2} \right).
  \end{align}
Now compute contractions of  the various terms in (\ref{taumn4}) with the polarization tensors. Taking into account that $\bar{h}_{MN}(q^M)$ in (\ref{ge_mom}) consists of
$u_M u_N$ and $\eta_{MN}$ terms, one can establish the rules:
\begin{itemize}
    \item the products of $ \delta^{\lambda}_{(M}
    \eta_{N)z}^{\vm{\lambda}}$ with $\va_{ij}^{MN}$ and
    $\va_{i}^{MN}$ vanish since  $e_i$ have no
    $z-$component;
   \item the combination $ k_{\lambda}\delta^{\lambda}_{(M}
    \eta_{N)z}^{\vm{\lambda}}$ has zero contractions with the polarization tensors
    $\va_{i\psi}^{MN}$ by virtue of orthogonality of $k^M$ and
    $e_i^M$ ;
    \item  the contractions of  $\Sigma_M^\mu
\Sigma_N^{\nu\vm{\mu}}  u_{\mu} u_{\nu}$ with all polarization tensors
are zero since $u_{\mu}$ has  only temporal non-vanishing
component, while the corresponding components of $e_i^M$ and $e_\psi^M$ are
zero;
    \item  the contraction $\Sigma_M^\mu
\Sigma_N^{\nu\vm{\mu}}  \eta_{\mu\nu} \va_{ij}^{MN}$ vanishes by
virtue of the mutual
    orthogonality of $e_i$ and $e_j$ in the brane spatial sector;
    \item  the term $\Sigma_M^\mu
\Sigma_N^{\nu\vm{\mu}}  \eta_{\mu\nu} \va_{i}^{MN}$ is zero
due to tracelessness in the brane spatial sector;
    \item the term $\Xi_{MN} \va_{i\psi}^{MN}$ vanishes since the vector
    $\underline{n}$ is unit and hence $\underline{n}\cdot
    \partial_{\theta_i}\underline{n}=0$;
    \item finally, the products of   $\,u_{ \lambda} u_{(M}^{\vphantom{\lambda}}\Sigma^{\lambda}_{N)}$
    from $\,h_{ \lambda\,(M}^{\vphantom{\lambda}}\Sigma^{\lambda}_{N)}$  with polarizations $\va_{ij}^{MN}$,
$\va_{i\psi}^{MN}$ and $\va_{i}^{MN}$ vanish since the vectors  entering them do not have both $t-$ and $z-$components.
\end{itemize}
Thus we see that the only non-zero polarization projection comes from $\varepsilon_z$, so one is left with the only scalar amplitude
$T_{z}(k)\equiv T_{MN}(k)\, \va_z^{MN}$. Moreover, similar considerations imply  that one can truncate the  relevant polarization tensor to
 \begin{align}\label{tip3a}
 \va_z^{MN}=\sqrt{\fr{D-3}{D-2}} \: e^M_\psi e^N_\psi \,.
 \end{align}

The   non-vanishing contractions are:
\begin{align}\label{tene}
 &\Sigma^{\mu}_{M}\Sigma^{\nu}_{N}
\eta_{\mu\nu}\va^{MN}_z= \sqrt{\fr{D-3}{D-2}}\, \sin^2\nhsp\psi\,, \qquad\quad
 k_{\la} \,
 \delta^{\lambda}_{(M} \eta_{N)z}^{\vm{\lambda}} \va^{MN}_z= -\sqrt{\frac{D-3}{D-2}}\: \omega\, \cos \psi \,\sin^2
 \nhsp \psi\, .
\end{align}
So collecting all the non-zero terms and integrating over $q$ we obtain:
 \begin{align}
 \un T_z(k)= - \sqrt{\fr{D-3}{D-2}}\fr{\vak^2 \mu m}{2}
 \fr{\gamma v\sin^2\nhsp\psi}{\omega^2+k_\bot^2 \gamma^2 v^2} \left[\frac{\omega^2 \cos\psi}{v\,(k_{\mu}k^{\kn\mu}+2i\ep k^0)}
\left(\gamma^2 v^2+\frac{1}{D-2}\right) +\gamma^2v^2-\fr{1}{D-2}
\right],\nn
\end{align}
where $k_\bot^2 \equiv \delta_{ij}k^i k^j$.
Taking into account  the on-shell condition $k_M k^M=0 $, this quantity can be rewritten as
 \begin{align}\label{T_br1}
 \un T_z(k)= - \sqrt{\fr{D-3}{D-2}}\fr{\vak^2 \mu m}{2 \omega^2}
 \fr{\gamma v\sin^2\nhsp\psi}{1+ \gamma^2 v^2\sin^2 \! \psi} \left[\frac{\cos \psi}{v\left[\cos^{2}\!\psi+2i\ep k^0\right]}
\left(\gamma^2 v^2+\fr{1}{D-2}\right) +\gamma^2v^2-\fr{1}{D-2}
\right] .
\end{align}
We note the infrared divergence of this amplitude at \mbox{$\omega\to 0$}, which is not surprising since our procedure did not take into account the finite depth of the piercing layer. We will deal with  this problem later on. Another interesting feature is that the amplitude remains non-zero in the limit \mbox{$v\to 0$}. This is related to branon excitation which takes place even with infinitesimal $v$, for more details see \cite{GaMeZ}. Also, one can see that the amplitude diverges as \mbox{$\psi\to \pi/2$}, i.e. along the DW. This divergence is another artefact of our approximation, to be dealt with later on.

The divergence of the brane amplitude along the wall is due to brane excitation $\Phi$.
If one  puts \mbox{$\Phi=0$} by hand (this may correspond to $\mathbb{Z}_2$-symmetric  braneworld models or to the case of two mirror particles impinging upon the wall), then
 \begin{align}\label{T_br1_reduc}
 \un T_z(k)\Bigr|_{\Phi=0}= - \sqrt{\fr{D-3}{D-2}}\fr{\vak^2 \mu \Eps}{2\omega^2}
 \fr{ v\sin^2\nhsp\psi}{1+ \gamma^2 v^2\sin^2\nhsp\psi} \left[\gamma^2v^2-\fr{1}{D-2}
\right] .
\end{align}
 Thereby in this case the amplitude does not blow up at $\psi=\pi/2$ and the angular distribution is finite.
 \subsection{The particle amplitude}
The first-order particle stress-tensor was found in the preceding section in the
coordinate representation  (\ref{T1MN}):
\begin{align} 
\un \,\bar{T}_{MN}(x)=  \frac{m}{2} \int   \left[ 4\un \dot z_{(M}
u_{N)} +\vak \lbr 4\,u^P h_{P(M}u_{N)} - {h} \, u_M u_N\rbr -
2\,u_M u_N \un {z}^P\pa_P \vp \right] \delta^D\!(x- u \tau)\,
d\tau\, \nn
\end{align}
with $z^M$ given by (\ref{cucu0}) and (\ref{acce2}). The corresponding amplitude in the momentum representation  reads:
\begin{align}\label{gg1}
\un \bar{T}_{MN}(k)=  \frac{m}{2} \int d\tau \, e^{i(ku)\tau}&
\left[  2\,i(k\un z)\, u_{M}u_{N}+4\, u_{(M} \un \dot{z}_{N)}-
\vak\,
 {h}(\tau)\, u_{M}u_{N}+4 \vak \, u^{P} u_{(M}h_{N)P}(\tau)\vp \right]  ,
\end{align}
where the brane gravitational field $h_{MN}(\tau)$ is given by (\ref{brgr}), restricted to the unperturbed trajectory:
 \begin{align} 
h_{MN}(\tau)=\frac{\vak \mu}2\left(\Xi_{MN}
-\frac{D-1}{D-2}\,\eta_{MN}\right)\gamma v |\tau|\, , \nn
 \end{align}
and $\un z^M(\tau)$ is given by (\ref{acce2}).
For $\un \bar{T}_{MN}(k)$  the  non-zero contribution gives only the polarization   $\va_z$, yielding the product $\un \bar{T}_{z}(k)$. Contracting (\ref{gg1}) with (\ref{tip3a}) and using the integrals
$$\int\limits_{-\infty}^{\infty} \e^{i \alpha \tau}|\tau|\,d\tau=
-\frac{2}{\alpha^2},\qquad  \int\limits_{-\infty}^{\infty} \e^{i
\alpha \tau}\tau^2 \, \sgn(\tau)\,d\tau=-\frac{4i}{\alpha^3},$$
one obtains
 \begin{align}\label{TmnpkLa}
\un \bar{T}_{z}(k)= - \left[\fr{D-3}{4(D-2)^3} \right]^{\!1/2} \frac{\vak^2 \mu m
v}{\gamma \om^2}\frac{  \left[(D-2)\kn\gamma^2 v^2-1\right] v\cos
\psi+2}{(1- v\cos \psi)^3}\,\sin^2\nhsp\psi\,.
\end{align}
This amplitude, apart from the infrared, has also the angular divergence at $\psi=0$ in the case of the massless particle $v=1$. This is the well-known  collinear  divergence encountered  in quantum perturbation theory for interacting massless particles. In classical theory this is the line divergence of the retarded potentials \cite{Lechner:2014kua}.
 \subsection{The  stress contribution}
The stress tensor is given by Eq.~(\ref{natag_0}). Substituting
$\Hh_{MN}=h_{MN}+\bar{h}_{MN}$ and keeping only the cross terms one
obtains a bi-linear form of $h$ and  $\bar{h}$. Anticipating zero contractions of $\eta_{MN}-$part with traceless polarization tensors, one can drop it from the beginning.
The Fourier transform of the product of two fields becomes convolution in the momentum representation presented by the integrals over the variable $q^M$. The following integrals are useful in such a computation:
 \begin{align}
\int \frac{\delta(qu) \,\delta^{D-1}(q^{\mu})}{q^2 (k-q)^2}\,d^D q =\frac{\gamma^3 v^3}{\rule{0cm}{0.9em} (ku)^3(\bar{k}u)}\,
 , \quad\qquad \int \frac{\delta(qu) \,\delta^{D-1}(q^{\mu})}{q^2 (k-q)^2}\,q_{M_1}\pp q_{M_n} \,d^D q
 =\left(\frac{ku}{ \gamma v} \right)^{n-3}\frac{ \delta_{M_{1}z} \pp\delta_{M_{n}z} }{ \rule{0cm}{0.9em} (\bar{k}u)} \nn
 \end{align}
(with $\bar{k}^{M}$ introduced in (\ref{bark})). After lengthy
but straightforward calculations one obtains the following expression:
 \begin{align}
  S_{MN}(k)= \fr{\vak^2 \mu m}
 {\gamma v\, \tilde{q}^2(k-\tilde{q})^2} & \left[ \Xi_{PQ}k^P k^Q\, u_{M}u_{N}
 +
 \left(\gamma^2 v^2-\fr{1}{D-2}\right)\tilde{q}_M \tilde{q}_N
 + (k u)^2\, \Xi_{MN}  + \right. \nn\\&
 \quad\left. + \left[(ku)+(\bar{k}u)\vp \right]\,\tilde{q}_{(M} u_{N)}-2\,(ku)\,
   k^{A} \hsp\Xi_{A(M} u_{N)}  \vph \right], \nn
 \end{align}
with $ \tilde{q}^M =(0,...,0, k^z-k^0\nhsp/v) .$ Here the longitudinal terms  proportional
to $k_{(M} B_{N)}$ (with any $D$-vector $B^{M}$)
were also omitted in view of transversality of polarization tensors.
On shell $k^2=0$ this expression reduces to
 \begin{align}
 S_{MN}(k)= \fr{\vak^2 \mu m \gamma^3 v^3}
 {  (ku)^3 \,(\bar{k}u) \rule{0cm}{0.9em}} & \left[ (k^z)^2\, u_{M}u_{N}
 +
 \left(\gamma^2 v^2-\fr{1}{D-2}\right)\tilde{q}_M \tilde{q}_N
 + (k u)^2 \,\Xi_{MN}  + \right. \nn\\&
 \;\left. +2\, \gamma\, k^0\,\tilde{q}_{(M} u_{N)} -2\,(ku)\,
   k^{A}\,\Xi_{A(M} u_{N)} \vph \right].\nn
 \end{align}
One can notice that the stress tensor contains the same tensor structures as  $\un \bar{T}_{MN}(k)$ and
$\un T_{MN}(k)$, so the only polarization tensor giving non-zero result will be again $\va_z^{MN}$. The following contractions can be easily found
\begin{align}
\tilde{q}_M e_{\psi}^M = -\frac{(ku)}{\gamma v}\,\sin \psi \,,
\qquad k^{A}\hsp \Xi_{A(M} u_{N)} \va_z^{MN} =
- \sqrt{\fr{D-3}{D-2}}\,\gamma v\, \omega \, \sin^2 \nhsp\psi\, \cos \psi\,,\nn
\end{align}
in addition to
(\ref{tene}). Using them, the projected stress-tensor amplitude can be presented as:
\begin{align}
S_z(k)=  \sqrt{\fr{D-3}{D-2}}\,
 \fr{\vak^2 \mu m\, v^3\,\sin^2\nhsp \psi}{ \gamma\hsp(k^0-v {k}^z)^3(k^0+v {k}^z)}\left[\gamma^2 v^2
 (k^z)^2- \fr{(ku)^2}{(D-2)\gamma^2v^2}
 \right]\,.\nn
\end{align}
Finally, using the on shell  parametrization $k=\omega \kn (1,\sin \psi\, \underline{n}, \cos \psi)$, one obtains:
\begin{align}\label{csL1}
S_z(k)= \sqrt{\fr{D-3}{D-2}}\,\fr{\vak^2 \mu m \,v}{\gamma\, \om^2}\,
 \fr{\sin^2\nhsp \psi}{(1-v \cos \nhsp \psi)^3(1+v \cos \psi)}\left[\gamma^2 v^4\cos^2\nhsp\psi
 - \fr{(1- v\cos \psi)^2}{D-2}
 \right] .
\end{align}
Here one  also observes both the infrared and  the angular  divergences.
\subsection{The destructive interference in the ultrarelativistic
limit}\label{DIf}
In the ultrarelativistic limit $v\to 1$ ($\gamma\to \infty$) both the particle and the stress amplitudes have similar behavior near the forward direction  $\psi\ll 1$ which could give the leading contribution to radiation (this follows from the
 integrals (\ref{jj0})). However,   keeping the common singular
factors and  expanding the rest as
\begin{align}\label{xi}
\sin \psi \approx \psi\, , \qquad 1-v \cos \psi \approx
\frac{\psi^2+\gamma^{-2}}{2}\,,
\end{align}
one finds  for large  $\gamma \gg1$
 \begin{align}\label{TmnpkLab}
\un \bar{T}_z(k) = -\frac{1}{2}\sqrt{\fr{D-3}{D-2}}
\,\frac{\vak^2 \mu \kn \Eps\, \sin^2\nhsp \psi}{ \om^2\,(1- v\cos
\psi)^3}\,\left(1+\mathcal{O}\kn(\gamma^{-2})\right)
\end{align}
and
\begin{align}\label{csL1b}
S_z(k)=  \frac{1}{2} \sqrt{\fr{D-3}{D-2}}\,
 \frac{\vak^2 \mu \kn \Eps\,\sin^2\nhsp  \psi}{ \om^2(1-v \cos \psi)^3}\,\left(1+\mathcal{O}\kn(\gamma^{-2})\right)  \, ,
\end{align}
where  $\mathcal{E} =m \gamma$ is the particle energy.  So in the leading in  $\gamma$ order,  these two amplitudes exactly cancel. This is manifestation of the \emph{destructive interference} which reflects the equivalence principle in the language of flat space, which was encountered in the bremsstrahlung problem for point particles
\cite{Khrip,Galtsov:1980ap, GKST-3, GKST-4, GKST-PLB}. After cancelation of the leading terms, the
sum of (\ref{TmnpkLa}) and
(\ref{csL1}) has two orders of gamma less than the each term separately.

On the other hand, the brane amplitude   (\ref{T_br1}) in the forward
 direction is approximated as
 \begin{align}\label{T_br2}
  \un T_z(k)\Bigl|_{\psi \ll 1} \approx - \sqrt{\fr{D-3}{D-2}}\fr{\vak^2 \mu \kn \Eps}{  \omega^2}
 \frac{\gamma^2 \sin^2\! \psi}{ 1+ \gamma^2\sin^2 \nhsp \psi  }
\end{align}
and thereby is of order of $ \mathcal{O} (\omega^{-2} \vak^2 \mu \Eps) $.  Comparing it with $(1-v \cos \psi)^{-2}\hm{=}
\mathcal{O}\kn(\gamma^4)$ one concludes that at \mbox{$\psi \approx 0$} the brane
contribution is always subleading.

\begin{figure}
\begin{center}
\includegraphics[angle=0,width=13cm]{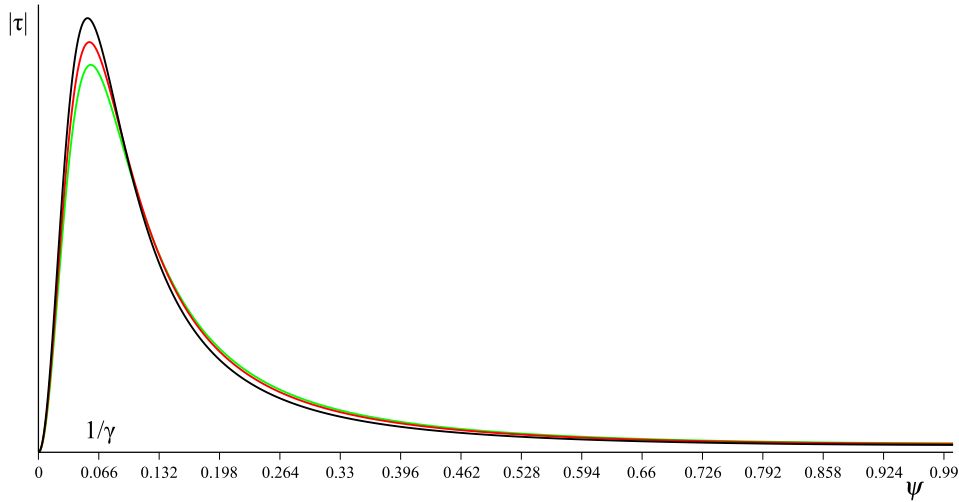}
\caption{The angular dependence of the  radiation amplitude for $\gamma=15$  in four (black), five (red) and six (green) spacetime dimensions (units
$\vak^2 \mu\kn  \Eps \omega^{-2}=1$ are understood).} \label{amplowdim}
\end{center}
\end{figure}

Thus in the small-angle region the main contribution still comes from the sum of  $\un \bar{T}(k^M)$ and $S(k^M)$.
Expanding these  with more accuracy and keeping the subleading
terms, one finds to the main order:
\begin{align} \label{tau_forw}
 \tau_z(k )\Bigl|_{\psi \ll 1} \approx -\frac{1}{8}
 \sqrt{\fr{D-3}{D-2}}
 \frac{\vak^2 \mu\kn  \Eps \,\sin^2\nhsp \psi}{
\hsp\gamma^2\,\om^2(1-v \cos \psi)^3 }\,\left[ \frac{D+2}{D-2} +  \gamma^2 \sin^2\nhsp\psi \vp\right]\,.
\end{align}
The total amplitude is peaked at \mbox{$\psi \sim 1/\gamma$} in any  dimensions\kn\footnote{More precisely, the position of maximum is $\psi = \Bigl(\frac{\sqrt{D^2+12}-4}{D-2}\Bigr)^{\akn 1/2}\!\bigl/\gamma$, plus small corrections.}, with the magnitude $\propto \gamma^2$. The qualitative picture in  $D=4,5,6$  is shown on the Fig.\,\ref{amplowdim}.

\begin{figure}
\begin{center}
\includegraphics[angle=0,width=13cm]{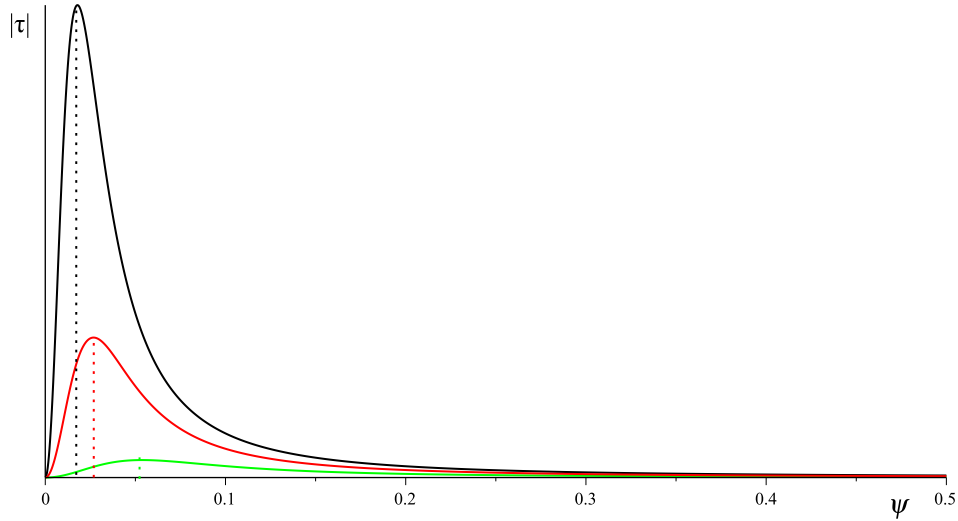}
\caption{The angular dependence of the  radiation amplitude in four dimensions for different values of Lorentz factor: $\gamma=15$  (green), $\gamma=30$  (red) and $\gamma=45$ (black) (units
$\vak^2 \mu\kn  \Eps \omega^{-2}=1$ are understood).} \label{ampldiffgamma}
\end{center}
\end{figure}

The dependence of the radiation amplitude  on the particle Lorentz factor in $D=4$ is shown on the Fig.\,\ref{ampldiffgamma}.

\subsection{The massless case}
The limiting case, corresponding to the piercing of DW by photons, is the case where mass tends to zero with fixed energy. The latter represents the photon's frequency in $\hbar=1$ units. The corresponding  change should be observable in the angular dependence of the radiation amplitudes. Indeed, the position of the maximum $\mathcal{O}(1/\gamma)$ for the ultrarelativistic particle goes to zero, while the height $\gamma^2$ blows up to infinity.

\begin{figure}
\begin{center}
\includegraphics[angle=0,width=13cm]{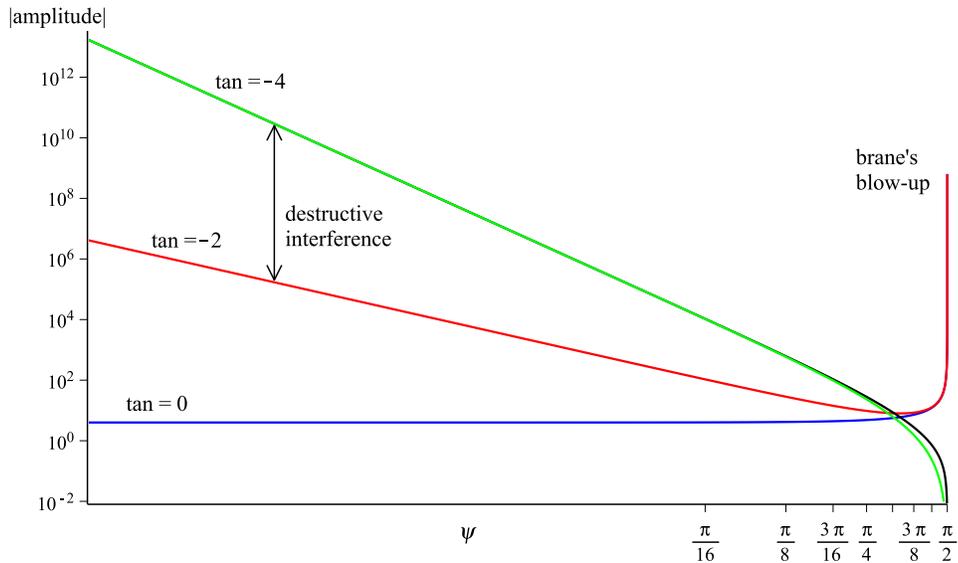}
\caption{Angular dependence of the  (absolute values) radiation amplitudes for photon:  particle  (black),   brane (blue) and stress (green) contributions and the total amplitude (red)  in doubly logarithmic mode  (in units
$C=1$).} \label{massless}
\end{center}
\end{figure}

Directly taking the limits of (\ref{T_br1}), (\ref{TmnpkLa}) and (\ref{csL1}) one gets
 \begin{align}\label{T_br1ml}
 \un T_z(k) & = -2C
  \left[\frac{\cos \psi}{\cos^{2}\!\psi+2i\ep k^0}
  +1
\right],&&   \un \bar{T}_{z}(k)  = - C\frac{ \cos
\psi \cos^2(\psi/2)}{\sin^4(\psi/2)}\,\nn \\
S_z(k)&= C\,
 \fr{\cos^2\nhsp\psi}{\sin^4(\psi/2)}\,,&&  C \equiv  \sqrt{\fr{D-3}{D-2}}\fr{\vak^2 \mu \Eps}{4\kn  \omega^2}\,,
\end{align}
with no dependence on the Lorentz factor. Here $\Eps$ stands for the photon frequency. One notices that for the small angles the brane contribution is regular, while both $\bar{T}_{z}(k)$ and $S_z(k)$
blow up as $\psi^{-4}$. Combining them, the total amplitude in the small-angle domain reads:
  \begin{align}\label{T_br1m2}
 \tau_z(k)  = -C\,\frac{\cos \psi}{\sin^2(\psi/2)}\,,
\end{align}
and thus blows up as $\psi^{-2}$ when $\psi$ approaches zero. Thus the destructive interference in this case consists in the diminishing  of the angular blow-up power
by two powers. The plot on Fig.\,\ref{massless} illustrates these observations.

Note that these curves are the same for any $D$, all $D$-dependence is contained only in the factor $C$.

\medskip

To summarize this section, we list the main results:
\begin{itemize}
\item
the radiation amplitude consists  of a single  polarization, responsible for emission into the bulk;
\item
the amplitude has the universal infrared divergence $\omega^{-2}$;
\item
in the ultra-relativistic case the radiation amplitude is peaked in the forward    direction, though the leading contributions of the particle and the stresses  mutually cancel due to destructive interference;
\item the brane amplitude blows up  along the DW and remains non-zero in the limit $v\to 0$.
\end{itemize}

 \section{The spectral and angular distribution of PGR}
From   Eq.~(\ref{epol}) one obtains  simple expression for the spectral-angular distribution of the total $\va_z$-polarized PGR:
\begin{align}\label{epol1}
\frac{dE_{\rm rad}}{d\omega d\Omega} =\frac{\vak^2}{4\left(2 \pi\right)^{D-1}}\,
 \omega^{D-2}
    \left|\tau_z(k)\right|^{2}.
\end{align}
 In view of the spherical symmetry on brane, we integrate over sphere  $S^{D-3}$ (all angles except $\psi$) obtaining for the total emitted energy
\begin{align}\label{epol1aa}
 E_{\rm rad}  =\frac{    \vak^2}{(4\pi)^{D/2}  \Gamma\!\left( \frac{D-2}{2}\right)}\,
 \!\int\limits_{\omega_{\min}}^{\omega_{\max}}\! d \omega\,  \omega^{D-2} \!\!\int\limits_{\psi_{\min}}^{\pi}\!d\psi \, \sin^{D-3}\psi
    \left|\tau_z( \omega, \psi)\right|^{2} .
\end{align}
The integrand typically is peaked along \mbox{$\psi=0$}, and for massive particles (finite $\gamma$) it is non-divergent there (actual integration is performed according to Appendix A). But in the massless limit it is divergent, so the cut-off at $\psi_{\min}$ is required. This cut-off depends on the particular physical problem which is supposed to replace our simplified model and may be either classical, or quantum. In what follows we will discuss this in more details.

Taking into account that all radiation amplitudes scale as $\omega^{-2}$ in the entire spectrum,
 the substitution of $\tau_z(k)$  into (\ref{epol1}) after integration over angles leads to the frequency distribution
\begin{align}\label{epol2}
\frac{dE_{\rm rad}}{d\omega} \propto      \vak^6 \kn \mu^2 m^2 \,
\omega^{D-6}.
\end{align}
This quantity is infrared-divergent in all dimensions less than six. This also is the consequence of the oversimplified nature of our model  which indicates the need of the infrared cut-off.  For  $D\geq 5$ an ultraviolet cut-off is also required.  These again can be classical or quantum, especially in the case of massless particles.  Recall that in the classical domain  we have  two intrinsic length parameters in the full non-linear theory ---  the inverse bulk curvature $k^{-1}$ and the gravitational radius associated with the particle energy $\mathcal{E}$, namely $$ r_{\mathcal{E}}\sim\left(\vak^2\mathcal{E}\right)^{1/{(D-3)}}\!\!.
$$ In addition, in more realistic DW models one encounters  other physical length parameters: the DW thickness $\delta$  and the finite longitudinal size of the DW. Finally, applicability of the perturbation theory requires  the distances in the bulk direction to be  restricted from above by the size of the piercing layer (\ref{vdl}) which is of the order of the bulk curvature radius for the photons, and is $O(v^2)$ for non-relativistic velocities.
 \subsection{The ultra-relativistic  case: beaming in the bulk direction}\label{URRad}
The dominant part of the total  radiation still can be beamed even if
the corresponding particle and stress contribution mutually cancel.
Substituting (\ref{tau_forw}) into (\ref{epol1aa}) one encounters
 the competing over the polar angle $\psi$: $V_6^{D+1}$, $\gamma^2 \,V_6^{D+3}$ and $\gamma^4 \,V_6^{D+5}$, as introduced and evaluated in the Appendix \ref{angleints}.
In the leading order one obtains
\begin{align}\label{epy1}
E_{\rm rad} = \frac{(D-3)\,\Gamma\!\left(\frac{D+6}{2} \right)\Gamma\!\left(\frac{6-D}{2} \right)}{30\,(D-2)^2(4\pi)^{D/2}\, \Gamma\!\left(D/2 \right)}\,(\vak^3 \mu  \Eps)^2  \, Q_D \,\gamma^{6-D} \,,
\end{align}
where  frequency factor  $\ds Q_D$ is
\begin{align}\label{frparr}
Q_D=\left\{%
\begin{array}{ll}
    \ds {1}/{\omega_{\min} }, & \hbox{$D=4$;} \\
    \ds \rule{0cm}{1.7em} \ln \frac{\omega_{\max}}{\omega_{\min}}, & \hbox{$D=5$;} \\
    \ds \rule{0cm}{1.5em}\omega_{\max}^{D-5}/(D-5), & \hbox{$D>5$.} \\
\end{array}%
\right.
\end{align}
The factor $\Gamma\!\left(\frac{6-D}{2} \right)$ indicates that for $D \geqslant 6 $ the formula (\ref{jj2q}) is irrelevant. Indeed, for $D=6$ the integral $V_{6}^{11}$  exhibits the non-beamed logarithmic behavior (\ref{jj3log}). The two remaining integrals are still   ''beamed''  (\ref{jj2q}), but the  power  of $\gamma$ in the denominator (\ref{tau_forw}) makes the forward-direction contribution to be of the same order as non-forward contributions which are more difficult to access analytically. For $D>6$ it is not hard to combine  contributions $V_6^{D+1}$, $ V_6^{D+3}$ and $ V_6^{D+5}$ (choosing the appropriate case out of (\ref{jj2q}), (\ref{jj3}) or (\ref{jj3log})), but the total forward radiation  is negligible due to the small phase volume of the forward-direction beaming cone $\psi \lesssim \gamma^{-1}$.
Thus it is natural to consider the cases $D=4$, $D=5$ and $D\geqslant 6$ for the ultrarelativistic  particle separately.

\medskip

 $\bullet \,\boldsymbol{D=4\kn.}$ The direct application of (\ref{epy1}) yields:
 \begin{align}\label{epy14D}
E_{\rm rad} = \frac{1}{80 \pi^2}   \frac{ (\varkappa_4^3 \mu  \Eps)^2 \gamma^{2}}{ \omega_{\min} } \,,
\end{align}
where we traded the particle mass in favor of the energy $\Eps=m\gamma$. Obviously this diverges in the massless limit  when $\gamma\to \infty$. This divergence  is a  consequence of the collinear divergence of the amplitude, which  requires  the angular cut-off $\psi_{\min}$:
 \begin{align}\label{epy14Das}
E_{\rm rad} = \frac{1}{80 \pi^2}   \frac{ (\varkappa_4^3 \mu  \Eps)^2  }{ \omega_{\min} \left(\psi_{\min}\right)^2} \,.
\end{align}
It is expected that the cut-off has quantum nature, like in the case of synchrotron radiation of massless particles \cite{SR}, which is beyond the scope of our treatment. Otherwise, one can think of the factor $(\psi_{\min})^{-1}$ as an effective maximal Lorentz factor $\gamma^*$.

\medskip

 $\bullet \,\boldsymbol{D=5\kn.}$ This case is  ''softer'' as containing a single power of $\gamma$: now from  (\ref{epy1}) we have:
 \begin{align}\label{epy15D}
E_{\rm rad} = \frac{7}{3\cdot 2^8 \pi^2}    \,(\varkappa_5^3 \mu  \Eps)^2 \gamma  \, \ln \frac{\omega_{\max}}{\omega_{\min}} \,.
\end{align}
In this case one can find reasonable cut-offs from applicability conditions of our approach.
 The cutoffs $\omega_{\min}$ and $\omega_{\max}$ come from the corresponding coordinate cutoffs discussed in the Appendix \ref{cutta}. Another restriction could come from non-infinite longitudinal size of physical DW, and one has to   combine them together. This stimulates us  to
 revisit
  the   1st-order deformation of the brane, given by $\Phi \equiv \Phi_{\ah} +\Phi_{\bh}$ \cite[eqns.\,5.14,\,5.17]{GaMeS1}, with
$$ \Phi_{\ah}= - \frac{\varkappa_5^2  \Eps}{8 \pi^2 r}\,\arctan \frac{r}{\gamma v t}\,, \qquad\qquad
\Phi_{\bh}=  - \frac{\varkappa_5^2  \Eps}{8 \pi  r}\, \theta(t)\, \theta(r-t)\,,$$
where $\theta(x)$ stands for the Heaviside step-function. The absolute value of both these  is  maximal for small $r$, hence the maximal $z-$direction brane deformation has to be taken at \mbox{$r=r_{\min}$} (for \mbox{$|t|>z_{\min}$}), or at \mbox{$r=0$}, \mbox{$t= z_{\min}$} in $\Phi_{\ah}$, yielding
$$ |\Phi|_{\max}= \frac{\varkappa_5^2  \Eps}{8 \pi^2 r_{\min} } \simeq \frac{3\re^2}{r_{\min}}\,,$$
where we take into account (\ref{co_app1}). Demanding for the brane deflection to be secured by the ''true'' minimal available-for-consideration
$z$-coordinate, in order to keep the  validity of the perturbation theory (what implies \mbox{$|\Phi|_{\max}=z_{\min}$)},  and taking into account the correlation between $z$ and $r$ for the branon wave
we conclude:
  \begin{align}\label{zmin5D}
r_{\min} \simeq z_{\min}= \mathcal{O}\kn(\re)\,,
\end{align}
since these exceed the ''previous'' minimal values \mbox{$r_{\min} =\re \gamma^{-1/(D-3)}$}, \mbox{$z_{\min} =\re \gamma^{-(D-2)/(D-3)}$} in (\ref{co_app1}). Thus
  \begin{align}\label{ommax5D}
 \omega_{\max}= \mathcal{O}\kn(1/\re)\,,
\end{align}
independent of the graviton direction.

Second, here we are basically consider the RSII setting where the brane is five-dimensionally infinite.  Let consider for the moment the case with infinite Lorentz factor within the RSII model. Note, the zeroth-order field in our model is Minkowskian everywhere. This fact preserves the conservation laws  \cite{GaMeS1, GaMeS2}, and the dynamics of particle and brane is self-consistent. But if we consider times $$t>L  \equiv z_{\max}$$ in
 RSII-setting, the free particle after the shock collision moves in the true RSII-metric which is exponentially decaying. Hence the corresponding branon wave, $\Phi_\bh$, propagating outward the piercing point, when reaching the value $r= L $ has to be deformed significantly\kn\footnote{Such a non-perturbative analysis lies beyond our goals here.}. In other words, we have to put
  \begin{align}\label{rmax5D}
r_{\max}= \mathcal{O}\kn(L)\,, \qquad\qquad  k^{r}_{\min}= \mathcal{O}\kn(1/L)\,.
\end{align}
The minimal value of frequency $\omega$ is determined by the two inverse length parameters: both longitudinal, $1/z_{\max}$, and transverse, $1/r_{\max}$. Noticing that the two latter coincide, we conclude
  \begin{align}\label{ommin5Dz}
 \omega_{\min}= \mathcal{O}\kn(1/L)\,.
\end{align}

Substituting  (\ref{ommax5D}) and (\ref{ommin5Dz}) into  (\ref{epy15D}) one obtains\kn\footnote{Despite the cutoffs are defined as approximate values of declared order, we have kept the numerical pre-factor due to the logarithm changes insignificantly with respect to the deviation of approximate $\omega_{\max}$ and  $\omega_{\min}$.}:
  \begin{align}\label{eradi}
E_{\rm rad} = \frac{7}{3\cdot 2^8 \pi^2}  \, (\varkappa_5^3 \mu  \Eps)^2  \gamma \, \ln \frac{L}{\re} \,.
\end{align}
Normalizing it by the particle energy, we have:
   \begin{align}\label{eff}
 {\epsilon} \equiv \frac{E_{\rm rad}}{\mathcal{E}} \simeq  \frac{7}{2^5 }   \,  \frac{\re^2 \gamma}{L^2}  \, \ln \frac{L}{\re} \,,
\end{align}
and we can consider two cases described in \cite{GaMeS3}: the resolution is that
 the existence of
$k^{r}_{\min}$ prevents the angle $\psi$ to approach zero:
$$\psi_{\min} \sim \arcsin\frac{k^{r}_{\min}}{\omega_{\max}} \simeq
\frac{\re}{L} \ll 1\,.
$$
Notice, the result (\ref{eradi}) is got after the $\psi$-integration from 0 to $\pi$,
whereas in the relativistic case we thus have to integrate the angular
distribution  (\ref{epol1}) from  $\psi_{\min} $.
Since the integrand is beamed  inside the cone $ 0< \psi\lesssim
1/\gamma $, the final result depends upon the relation between $
1/\gamma$ and $ \psi_{\min} $. Namely, expressing the radiation efficiency (\ref{eradi}) in terms of
 inverse minimal emission angle, it estimates as
 \begin{align}\label{tyu1}
 {\epsilon} \sim   \frac{\re}{ L} \,  \ln\frac{L}{\re}  \,,
\end{align}
or $\ln \gamma_{\ast}/\gamma_{\ast}$  in  terms of effective Lorentz factor $\gamma_{\ast}\equiv \psi_{\min}^{-1}$.
Since the function $ \ln x/x$ does not exceed 1 for $ x>1$,
there is no efficiency catastrophe in our model. In fact, we assume \mbox{$L/\re  \simeq \gamma_{\ast} \gg 1$}, so  \mbox{$ {\epsilon}  \ll 1$}.

\medskip

 $\bullet \,\boldsymbol{D\geqslant 6\kn.}$ As was deduced above, the most of the angular contribution is taken from the angles of order $\mathcal{O}\kn(1)$. Meanwhile, in six dimensions both the ''beaming''\xspace region $\psi \lesssim \mathcal{O}\kn(\gamma^{-1})$ and $\psi \gtrsim \mathcal{O}\kn(\gamma^{-1})$ contribute on equal footing, that implies for the local values of the angular-distribution curve at $\psi \simeq \mathcal{O}\kn(\gamma^{-1})$ to be in $\gamma$  times greater than the same one at $\psi \simeq \mathcal{O}\kn(1)$. The plot on Fig.\,\ref{baldas} confirms this conclusion. Thus we have the ''local'' beaming, with the total contribution as
   \begin{align}\label{energy6D}
 E_{\rm rad} =\frac{6}{(4\pi)^5} (\varkappa_6^3 \mu\kn  \Eps)^2 \left(\ln 2\gamma - \frac{107}{60}\right)\omega_{\max}\,, \qquad\quad D=6\kn.
\end{align}

For $D>6$ the $\gamma \gg 1$ limit has to be applied to the region  $\psi=\mathcal{O}\kn(1)$. Doing this, one obtains the estimate
   \begin{align}\label{energy7D}
\tau_z(k )\Bigl|_{ \psi = \mathcal{O}(1)}  \sim
 \frac{\vak^2 \mu\kn  \Eps}{\omega^2} f(\psi)\,, \qquad\quad E_{\rm rad} \sim  (\vak^3 \mu\kn  \Eps)^2 Q_D \,,
\end{align}
where $f(\psi)$ denotes some function of order $\mathcal{O}\kn(1)$ with no characteristic dependence on \mbox{$\gamma\gg 1$} inside.

\subsection{Non-relativistic case: the brane contribution}
\begin{figure}
\begin{center}
\includegraphics[angle=0,width=13cm]{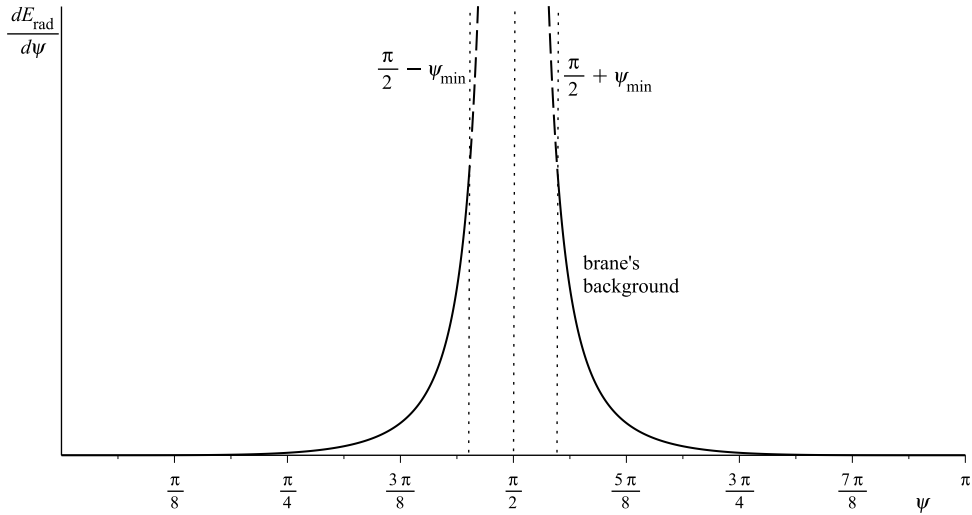}
\caption{Angular distribution of the emitted radiation at zero-speed for $D=4$.} \label{branebg}
\end{center}
\end{figure}

As it was mentioned above, the brane contribution survives in the limit $v\to 0$ (\ref{T_br1}). In this case the angular distribution is symmetric with respect to the brane's plane.  Thus we deal with the pure brane configuration  of waves, the corresponding plot is shown on the Fig.\,\ref{branebg}.

However, the term $1/k_z$ in ${\un T_z(k)}$ is presented in the generic case $v>0$, thus one can expect the blow up of the curve
 of the total emission angular distribution  at angle $\psi$ close to $\pi/2$, i.e. when the graviton is emitted tangent to the brane:
from (\ref{T_br1}) we have now:
  \begin{align}\label{T_br1_tang}
\tau_z(k)\Bigr|_{\psi \approx \pi/2}= - \sqrt{\fr{D-3}{D-2}}\fr{\vak^2 \mu m}{2\kn \omega^2 \cos\psi}
 \fr{\gamma}{1+ \gamma^2 v^2}
\left(\gamma^2 v^2+\fr{1}{D-2}\right).
\end{align}
Introducing the complement $\psi'\equiv \pi/2-\psi$ and estimating, we conclude:
  \begin{align}
\tau_z(k)\Bigr|_{\psi \approx \pi/2} \simeq - \fr{\vak^2 \mu \Eps}{\omega^2 \psi'}\,.\nn
\end{align}
 This indicates to the imposing of a cutoff on the $k^z$ and $\psi'$: indeed, from the maximal $z_{\max}$ we can deduce:
$$ k^z_{\min} = \mathcal{O}\kn(1/L)\,, \qquad \quad  \psi'_{\min}  =\psi_{\min}\,. $$
Integration of (\ref{T_br1_tang}) yields the ''brane''\xspace contribution to the spectrum:
 \begin{align}\label{T_br1yf}
E_{\rm br} \simeq  (\vak^3 \mu \Eps)^2
 \frac{Q_{D+1}}{k^{z}_{\min}} \,,
\end{align}
with $Q_n$ introduced in (\ref{frparr}).
However, in the non-relativistic case the maximal $z$-size of applicability of our approach is $z<z_l(v)$ which goes to zero as $v^2$ for small velocities. So in the non-relativistic case one could do better considering particle motion in the exact brane background, we leave this for the future work.

 \medskip

\begin{figure}
\begin{center}
\includegraphics[angle=0,width=13cm]{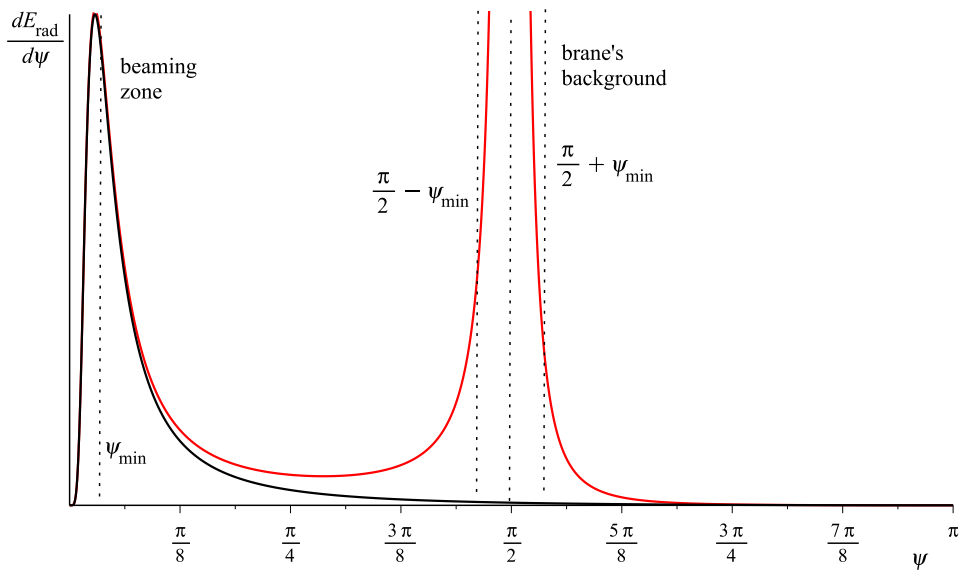}
\caption{The total angular distribution of the emitted radiation for full source (red) and reduced one, with no brane excitation (black) for $D=5, \gamma=15$.} \label{PhiornotPhi}
\end{center}
\end{figure}

 \textbf{$\mathbb{Z}_2$-symmetric case.} When both half-spaces of the brane hyperplane are equivalent, the brane excitation is absent:
 $$\Phi=0\,.$$
 Thus the brane contribution to the total radiation amplitude is given by\kn\footnote{The amplitude, emitted energy and radiation efficiency, corresponding to this mirror case, will be denoted with star}:
 \begin{align}\label{T_br1_reduc2}
  \un T^{*}_z(k) \equiv \un T_z(k)\Bigr|_{\Phi=0}= - \sqrt{\fr{D-3}{D-2}}\fr{\vak^2 \mu \Eps}{2\omega^2}
 \fr{ v\sin^2\nhsp\psi}{1+ \gamma^2 v^2\sin^2\nhsp\psi} \left[\gamma^2v^2-\fr{1}{D-2}
\right] .
\end{align}
 Thereby in this case the amplitude does not blow up at $\psi=\pi/2$ and the angular distribution is finite. The characteristic plot
 with/without counting of $\Phi$ is shown on Fig.\,\ref{PhiornotPhi}. One sees that the reduction of $\Phi$ eliminates the infinite brane-motivated background at $\psi\sim \pi/2$.

\begin{figure}
\begin{center}
\includegraphics[angle=0,width=13cm]{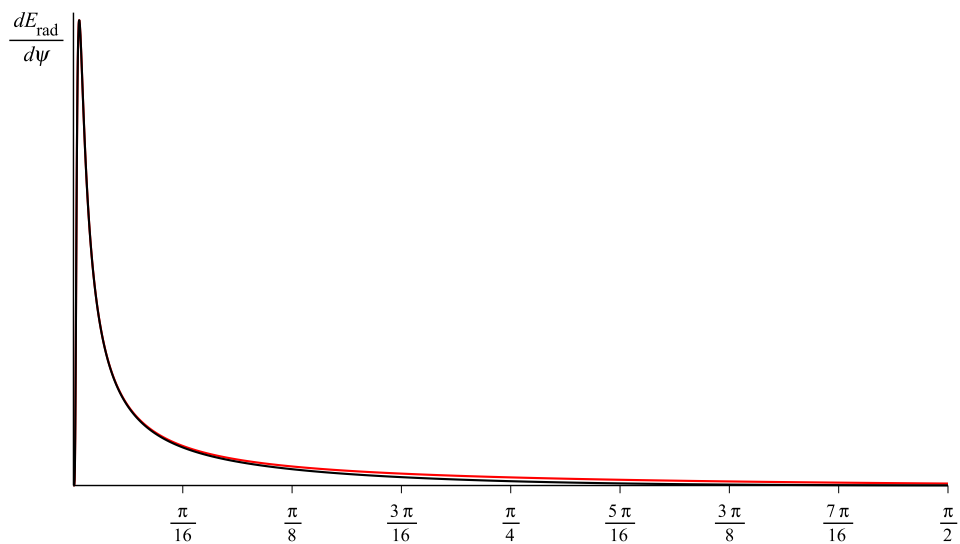}
\caption{The reduced angular distribution of the emitted radiation (red) versus the approximation by formula (\ref{tau_forw}) (black) for $D=6, \gamma=200$. For higher values of $\gamma$ two curves are more close to each other.} \label{baldas}
\end{center}
\end{figure}

The total angular distribution may be integrated for \mbox{$D\geqslant 6$} in the ultra-relativistic case\footnote{Remarkably, the validity of the low-angle approximation is extended to the angle region of order $\mathcal{O}\kn(1)$, as it shown on Fig.\,\ref{baldas}}.
For \mbox{$\gamma \gg 1$} one obtains:
  \begin{align}\label{Tau_reduc_highangl}
  \tau^{*}_z(k)\Bigr|_{\psi\sim 1,\Phi=0, \gamma \gg 1}=   \sqrt{\fr{D-3}{D-2}}\fr{\vak^2 \mu \Eps}{\omega^2}
 \fr{1/2}{1-v\cos \psi}\,.
\end{align}

\begin{figure}
\begin{center}
\includegraphics[angle=0,width=11cm]{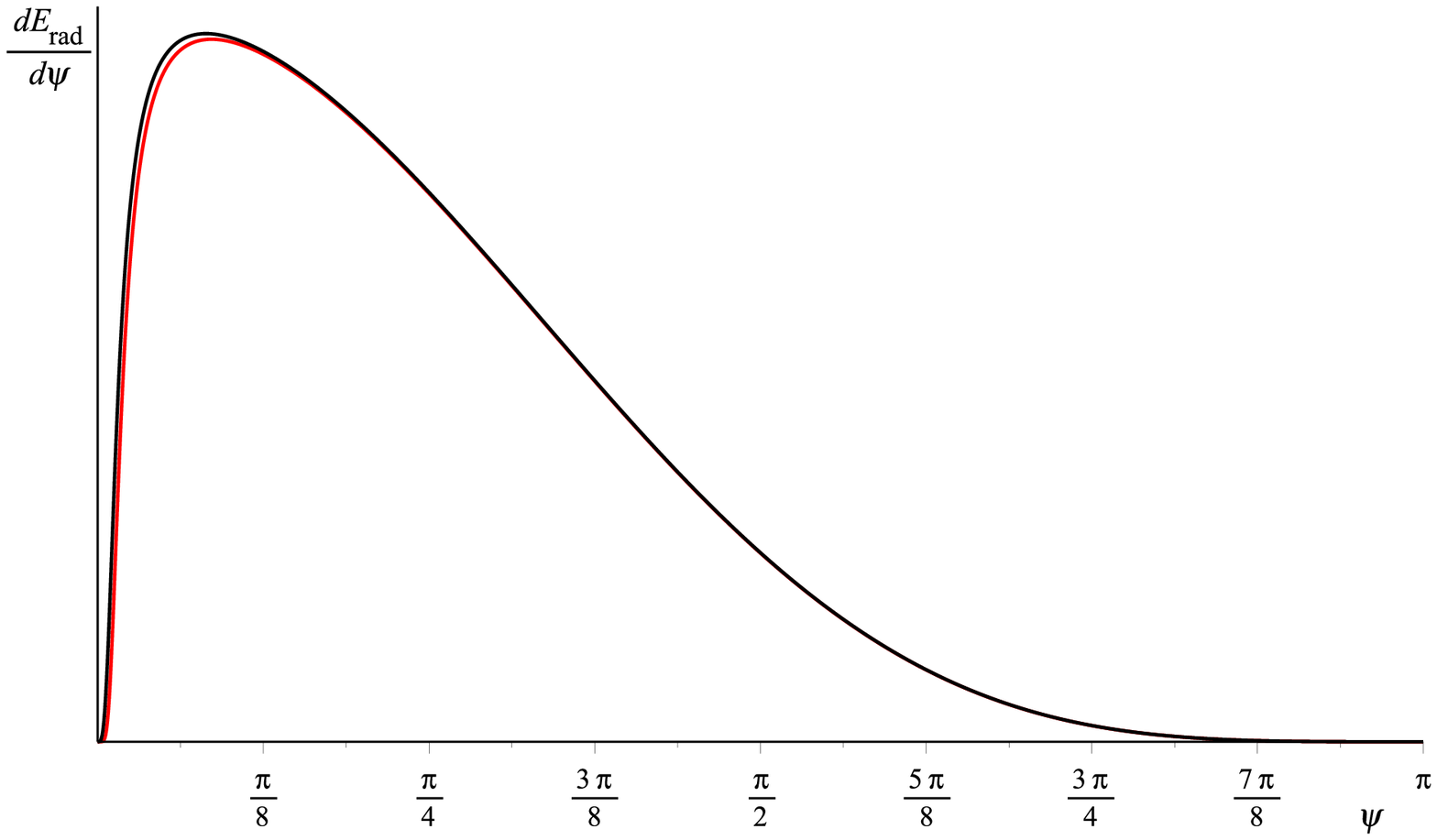}
\caption{The reduced angular distribution of the emitted radiation (red) versus the approximation by formula (\ref{Tau_reduc_highangl}) (black) for $D=7, \gamma=30$. For higher values of $\gamma$ two curves are more close to each other.} \label{burratino}
\end{center}
\end{figure}

The plot illustrating this approximation, is given on Fig.\,\ref{burratino}. Squaring of (\ref{Tau_reduc_highangl}) and substituting it into (\ref{epol1aa}) leads to the integrals $V_2^{D-3}$ which for \mbox{$D>6$} turn out to be ''non-beamed'' and for \mbox{$D=6$} logarithmic\kn\footnote{The result reproduces (\ref{energy6D}) with logarithmic precision.}, thus taking help of (\ref{jj3}) this yields:
  \begin{align}\label{E_reduc_highangl}
 E^{*}_{\rm rad} =\frac{1}{2^7\pi^{D/2}} \frac{(D-3)\, \Gamma\!
 \left(\frac{D-6}{2}\right)}{(D-2)\,(D-5)\, \Gamma
 (D-4)}(\varkappa^3 \mu\kn  \Eps)^2 \omega_{\max}^{D-5}  \qquad\quad  D\geqslant 7\,,
\end{align}
with $\omega_{\max}$ discussed above (\ref{ommax5D}).

\begin{figure}
\begin{center}\noindent
\includegraphics[angle=0,height=5.5cm]{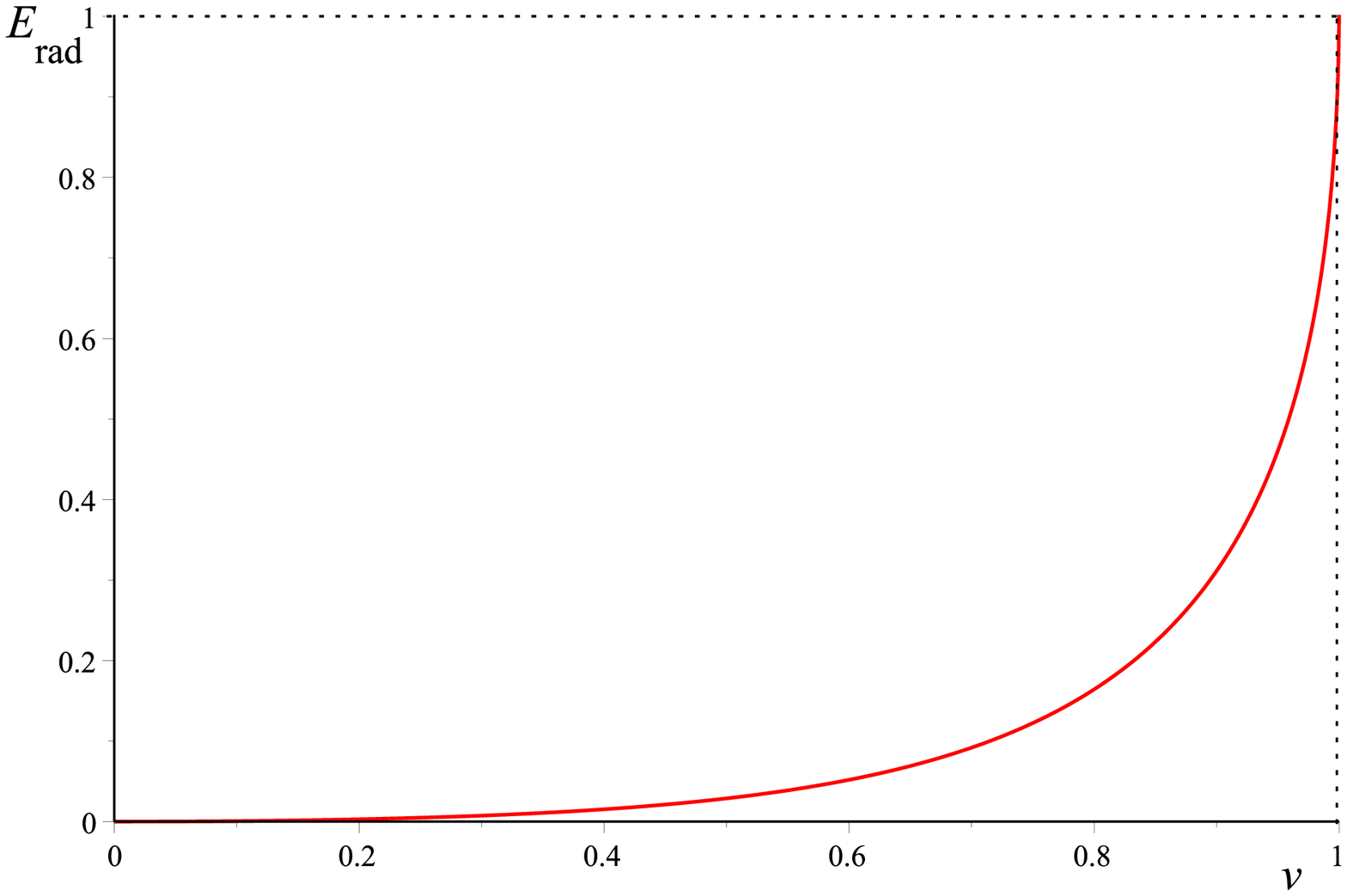}  \hfill \raisebox{-1.5pt}{\includegraphics[angle=0,height=5.5cm]{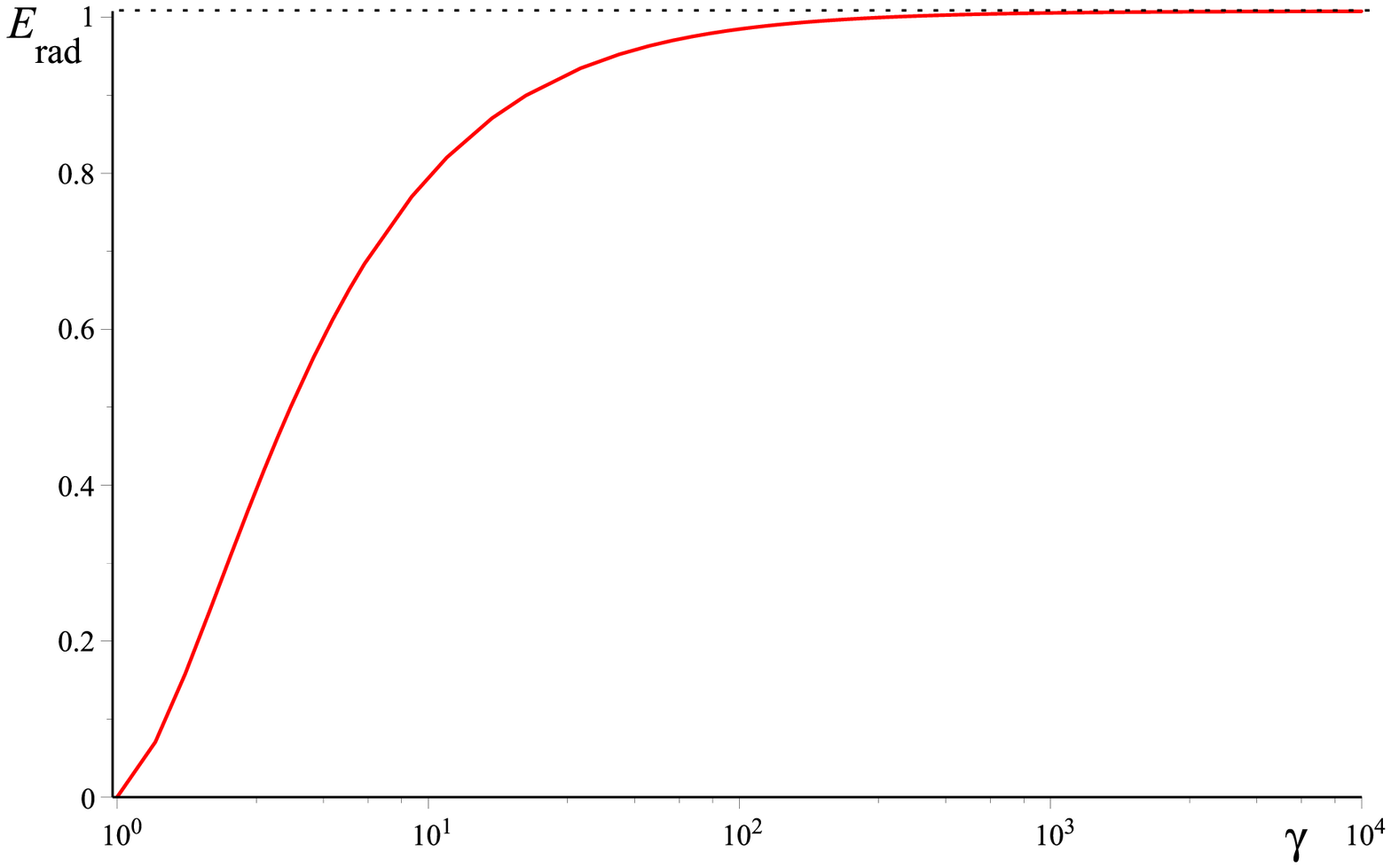}}\\[-10pt]
\hphantom{.} \hfill \emph{a})  \hfill \hfill \hfill \emph{b})  \hfill \hphantom{.}
\caption{The   emitted radiation against the speed of particle (\emph{a}) and Lorentz factor (in logarithmic mode) (\emph{b})  in $D=7$ dimensions, normalized by the maximal value given by eq.\,(\ref{E_reduc_highangl}).} \label{total7D}
\end{center}
\end{figure}

The formula (\ref{E_reduc_highangl}) obtained for an ultrarelativistic particle, does not contain ``free'' Lorentz factor. Also, the angular cut-off
 at small angles is not required.
 Thus  in higher dimensions there is no strong enhancement of the emitted radiation in the ultrarelativistic case. Omitting the brane's blow-up, the dependence of the emitted energy upon the speed of piercing particle is shown on Fig.\,\ref{total7D}.

The radiation efficiency is estimated as:
$$ \epsilon^{*} \sim   \Bigl(\frac{r_{\Eps}}{L}\Bigr)^{\akn 2}  \ln \frac{L}{r_{\Eps}}, \; D=6 \qquad\qquad
 \epsilon^{*} \sim \Bigl( \frac{r_{\Eps}}{L} \Bigr)^{\akn 2}, \;D>6\,. $$

To summarize, the total efficiency of radiation can conveniently be expressed through the ratio of the beamed part  to the brane contribution \mbox{$\chi \equiv E^*_{\rm rad}/E_{\rm br}$} giving
$$ \epsilon \simeq \epsilon^* \theta(\chi-1) + \frac{\epsilon^*}{\chi}\,\theta(1-\chi)\,,$$
where $\theta(x)$ is the Heaviside step-function. In four and five dimensions the beamed radiation dominates, while in higher
dimensions the most of radiation is emitted at large angles.

\section{Conclusions}
In this paper we investigated new mechanism of gravitational radiation from DWs, called PGR, due to their collisions with surrounding particles which perforate DWs and pass through. Within our model the DW-particle interaction was assumed to be purely gravitational and small, so the perturbative scheme on the Minkowski background is applicable. The gravitational force in this system is repulsive, so the particle must have the momentum transverse to the wall large enough in order to overcome repulsion and to perforate the wall. In plasma with some velocity distribution such particles form a layer whose size depends  on their velocity; this size becomes large in the ultrarelativistic limit and for massless particles.

In the perturbation theory, gravitational radiation arises in the second, post-linear order. This is similar to perturbative treatment of gravitational bremsstrahlung under relativistic collisions of point particles, but in our case situation is more complicated since the wall is an extended object with an intrinsic dynamics due to tension. Another difference is that the force between the particle and the wall does not fall off with distance, so there are no asymptotically free states. So to calculate gravitational radiation in such a  collision we had to resolve some conceptual and technical problems. One problem is failure of the traditional theory of radiation based on the notion of the wave zone which is absent in our case. So we had  to reconsider  formula for gravitational radiation without recurring to asymptotic conditions.
The second problem is that, due to absence of free asymptotic states, perturbative description of gravity is restricted to certain distances around the wall, while the formal expansions in terms  of  gravitational coupling constant  require  considering in zero order an infinite motion of the free particle. Also, the point particle is not a good approximation itself, since any mass has an associated gravitational radius.
Similarly, physical DW has a finite width, while we used the Nambu-Goto action. These oversimplifications allowed us to compute the radiation amplitudes analytically at the  expense  of the infrared and the ultraviolet  divergences of the spectra, as well as (collinear) divergences in the angular distribution in the limit of massless particle. So we had to perform an additional analysis to motivate the choice of cutoffs needed to extract finite answers from the perturbation theory.

Due to the symmetry of the problem, the emitted radiation is polarized  in single tensorial state, responsible for the bulk emission of GW.
The radiation amplitude consists of the  contribution due  to the particle, to the DW and to gravitational stresses. The total radiation amplitude is factorized into the product of the frequency part and the angular part depending upon the single angle.
The frequency factor $\omega^{-2}$ has an infrared blow-up, proper for the particles collisions \cite{Weinberg} in general, though stronger.   The angular part has divergence corresponding to gravitons emitted along  the brane which is absent in the case of two mirror particles.  In the non-relativistic limit  the particle and stress contributions  vanish, so only the brane contribution remains.
For an ultrarelativistic particle  the radiation amplitude in the  forward    direction is damped by two powers of $\gamma$ with respect to the particle term only due to   destructive interference with the stress tensor contribution.
Despite this, the radiation amplitude has a peak at \mbox{$\psi \sim 1/\gamma$} (with the same order of half-width), proper to  radiation from fast particles (bremsstrahlung), with magnitude $\sim \gamma^2$.

The angular-frequency distribution of the emitted radiation, obtained after squaring the total amplitude and adding the dimension-dependent phase-volume factor, have the following features. The frequency distribution scales as $\sim \omega^{D-6}$ in the entire spectrum and therefore has infrared divergence in four dimensions, the ultraviolet divergence in higher dimensions, and both of them for \mbox{$D=5$}. All of them are removed by the appropriate cutoffs which may depend upon the particular DW model.
The amplitude peak, discussed above, causes the beaming of the emitted gravitons, with characteristic cone angle $1/\gamma$, as it common for fast particles.
Such a beaming is realized in four and five dimensions, that reflects the corresponding enhancement of the emitted energy as gamma-factors in numerator. In the massless limit one uses the cutoff related with the applicability of linearized fields.
However, in higher dimensions the dominant  region of angular distribution shifts to the large angles, due to volume factor of the phase space. In this case
the radiation tangent-to-brane becomes dominant even with cutoff imposing. Thus in higher dimensions there is no strong enhancement of the emitted energy for relativistic/ultrarelativistic velocities, as well as no beaming of the emitted waves.

In the massless limit the radiation amplitude has no peak and blows up as $\psi$ approaches zero. However, the resulting angular amplitude scales as $\sin^{D-7}\!\psi$ and  regular for $D>6$. The cutoff imposing solves this divergence at \mbox{$D=4\,...\,6$} in the same manner like for ultrarelativistic case with finite Lorentz factor exceeding the effective ''cutoff'' factor $\gamma_*$.
In the non-relativistic limit the radiation due to the brane deflection dominates.  Omitting the brane's influence, the radiation is roughly isotropic.
The radiation efficiency depends on the braneworld  model and dimensionality. In higher dimensions with infinite-sized brane it is found to be of order $\ln \gamma / \gamma$ or less, where for very high Lorentz factor one has to substitute it by effective  \mbox{$\gamma_{*}=L/r_{\Eps} \gg 1$}.

\section*{Acknowledgments}
The work was supported by the Russian
Foundation of Fundamental Research under the project 17-02-01299a. DG also acknowledges the Russian Government Program of Competitive Growth of the Kazan Federal University.

\appendix
\section{Angular integration} \label{angleints}
Here we compute  the angular integrals of the
generic form
\begin{align}\label{jj0}
V_{m}^n=\int\limits_0^{\pi}\frac{\sin^n\nhsp
\psi}{\left(1-v\cos\psi\right)^m}\,d\psi
\end{align}
with integers $m, n$, and estimate them in the leading order in $\gamma \gg 1$. Making use of the formula \cite{Proudn}
\begin{align}\label{jj1pru}
 & \int\limits_0^\pi\frac{\sin^{2\nu-1}{\theta}}{\left(a-b\cos{\theta}\right)^{m}}\,d \theta=
\bigl(  {2}/{b}\bigr)^{\nu-1/2} \sqrt{\pi}\,
\Gamma{(\nu)}
 \left(a^2-b^2\right)^{\frac{2(\nu - m
 ) -1}{4} }P^{1/2-\nu}_{\nu-m-1/2}\!\left(\frac{a}{\sqrt{a^2-b^2}}\right),
\end{align}
valid for any real $a>|b|$, and $ \mathrm{Re} \, {\nu}>0$, we express the result in terms of the associated Legendre function
of the first kind $P^\mu_\nu(z)$. In our case  $a=1, b=v,$ so:
\begin{align}\label{jj1s}
  V_{m}^{n}=
 \left({2}/{v}\right)^{n/2}   \sqrt{\pi} \,\Gamma
 \Bigl(\frac{n+1}{2}\Bigr)
 \gamma^{-n/2+ m}P^{-n/2}_{n/2-m} (\gamma ).
\end{align}
In non-relativistic limit \mbox{$v \ll 1$} we start directly from
(\ref{jj0}) and expand it in powers of $v$. Integration over
$\psi$ gives
\begin{align}\label{nonrel}
  V_{m}^{n}=\int\limits_0^{\pi} \bigl(1+ m\,v\cos\psi\bigr)\, \sin^n\nhsp
\psi \,d\psi+\mathcal{O}\hsp(v^2)=\frac{\sqrt{\pi}\:\Gamma\!
 \left(\frac{n+1}{2}\right)}{\Gamma\!
 \left(\frac{n+2}{2}\right)}+\mathcal{O}\hsp(v^2)\,.
\end{align}
For $\gamma \gg 1,$ one can use the asymptotic formula  \cite{GR}:
\begin{align}
\label{jj2}
 P^{\lambda}_{\rho}(\gamma)=\left(\frac{\Gamma(-\rho -
1/2)} {2^{\rho+1}\sqrt{\pi}\,\Gamma(-\rho -\lambda)}\,
\gamma^{-\rho-1} +\frac{2^{\rho} \Gamma(\rho+1/2)}
{\sqrt{\pi}\,\Gamma(\rho -\lambda +1)} \, \gamma^{\kn\rho} \right)\!
\left(1+O\kn(\gamma^{-2})\right).
\end{align}

For $2m>n+1$ the  leading power of $\gamma$ comes from the
first term in  parenthesis of (\ref{jj2}), thus retaining it one
finds
\begin{align}\label{jj2q}
V_{m}^{n}=  \frac{   2^{m-1}   \Gamma\!
 \left(\frac{n+1}{2}\right)
 \Gamma \!\left(m-\frac{n+1}{2}\right)}{ \Gamma(m)} \,\gamma^{2m -n-1} \,.
\end{align}
Being applied to the computation of radiation flux, this type
corresponds to beamed emission inside the spatial cone with
characteristic angle of order \mbox{$\mathcal{O}\kn(\gamma^{-1})$}. Beyond
this cone the integrand in (\ref{jj0}) decreases rapidly. An alternative derivation of this answer can be found in appendix of \cite{Giannis}.

For $2m<n+1$ the behavior of Legendre function is governed by the
second term in  parenthesis of (\ref{jj2}); thus   one obtains
\begin{align}\label{jj3}
V_{m}^{n}=
           \frac{2^{n-m}   \Gamma\!\left(\frac{n+1}{2}\right)
           \Gamma\!\left(\frac{n+1}{2}-m\right)} { \Gamma(n -m+1
           )}\, .
\end{align}

In the borderline case
  $2m=n+1$ the behavior of the integral is logarithmic. Indeed,
 inserting the expansion (\ref{xi}) into (\ref{jj0}) one
 integrates $1/\psi$ from  ${\mathcal{O}\hsp(\gamma^{-1})}$ to
 ${\mathcal{O}\hsp(1)}$. Both terms in (\ref{jj2}) become actual, hence representing \mbox{$n=2m-1+\epsilon$}, we take the well-defined limit \mbox{$\epsilon\to 0$}, to obtain an asymptotic in this transition case:
\begin{align}\label{jj3log}
V_{m}^{2m-1}=    2^{m} \nhsp \left[ \ln 2\gamma
 - H_{m-1} \vp\right]+\mathcal{O}\hsp(\gamma^{-2})\, ,
\end{align}
where $H_{n}=\sum\limits_{k=1}^n \! k^{-1}$ stands for the $n$-th harmonic number. It has to be pointed out that for contemporarily reasonable values
of the Lorentz factor $\gamma\gg 1$, the $\mathcal{O}\hsp(1)-$term
is comparable with $\mathcal{O}\hsp(\ln\gamma)-$term. This justifies the
presence of non-logarithmic term here.
In the special cases of interest here, with $m=2$ and 6, integrals of this
type are given by
\begin{align}\label{jj3log6}
 V_{2}^{3}= 4 \bigl(\ln 2 \gamma -1\bigr),   \qquad\quad    V_{6}^{11}=   64 \! \left( \ln 2\gamma
 -  \frac{137}{60} \vp\right) .
\end{align}

Furthermore, the generalization of  angular integrals of the
  form
\begin{align}
V_{m_1,m_2}^n=\int\limits_0^{\pi}\frac{\sin^n\nhsp
\psi}{\left(1-v\cos\psi\right)^{m_1} \! \left(1+v\cos\psi\right)^{m_2}}\,d\psi\nn
\end{align}
\emph{in UR limit} can be reduced into the case of single $V_m^n$ if $m \equiv \max\{ m_1,m_2\} >(n+1)/2$. In this case the domination domain
is $0\pp \mathcal{O}\hsp(1/\gamma)$ if $m=m_1$, $ \pi- \mathcal{O}\hsp(1/\gamma)\pp \pi$ if $m=m_2$ and both of them if $m=m_1=m_2$.
The leading in $\gamma$ order thereby is given by
\begin{align}\label{jj0sd1}
V_{m_1,m_2}^n=\left\{%
\begin{array}{ll}
    2^{-m_2} V_{m_1}^n, & \hbox{$m_1>m_2$;} \\
   \rule{0cm}{1.3em} 2^{-m+1} V_{m}^n , & \hbox{$m_1=m_2$;} \\
   \rule{0cm}{1.3em}  2^{-m_1} V_{m_2}^n, & \hbox{$m_1<m_2$.} \\
\end{array}%
\right.
\end{align}
where $V_{m_1}^n$ and $V_{m_2}^n$  are given by (\ref{jj2q}). For arbitrary $v$ or if $m \leqslant (n+1)/2$,
the integrals of this type can be computed numerically.

\section{Applicability of the perturbation theory} \label{cutta}

The linearized field generated by the DW approximately coincides with the full non-linear solution iff the bulk distance is small with respect to the brane-curvature radius, the latter is given by
$$L\equiv (\vak^2 \mu)^{-1}.$$
In order to justify the correspondence of the linearized field generated by the particle one has to take into account the motion. In the particle rest frame the metric generated by the particle, is the Schwarzschild -- Tangherlini one: in the isotropic coordinates $(\tilde{t},\tilde{x}^{1}, \pp , \tilde{x}^{D-1})$ it is given by
\begin{equation}\label{STiso}
\mathrm{d}s^2  = \frac{\ds \left[1-(\varrho_g/\varrho)^{D-3}\kn
\right]^2}{\left[1+(\varrho_g/\varrho)^{D-3}\kn\right]^2}\, \mathrm{d} \tilde{t}^2
-
\biggl[1+\Bigl(\frac{\varrho_g}{\varrho}\Bigr)^{\!D-3}\biggr]^{\frac{4}{D-3}}\!\akn
\akn \delta_{ij}\,\mathrm{d}\tilde{x}^{i} \mathrm{d}\tilde{x}^{j},
\end{equation}
where $\varrho\equiv \left(\delta_{ ij }\, \tilde{x}^{i}
 \tilde{x}^{j}\right)^{1/2}$ and $i,j$ run from 1 to $D-1$. Here $\varrho>\varrho_g$ and  $\varrho_g$ stands for the \emph{isotropic-coordinate} gravitational radius, which is given by
\begin{equation}
 \varrho_g
=\frac{1}{\sqrt{\pi}}\biggl[\frac{2 \,
\Gamma\!\left(\frac{D-1}{2}\right)G m}{D-2}\biggr]^{\frac{1}{D-3}}\!\!.
\end{equation}
Now we transform (\ref{STiso}) to our Lab frame which is the unexcited-brane rest frame, with coordinates ($t, {x}^{1}, \pp ,  {x}^{D-2}, z$) used in the main text. Applying the Lorentz boost in $z-$direction $t=\gamma\kn(\tilde{t}+v\tilde{x}^{D-1})$, $z=\gamma\kn(\tilde{x}^{D-1}+v\tilde{t})$, $x^i= \tilde{x}^{i}$,
and introducing the dimensionless factor $$A \equiv \Bigl(\frac{\varrho_g}{\varrho}\Bigr)^{D-3} =\frac{2 mG\,
\Gamma\!\left(\frac{D-1}{2}\right)}{(D-2) (\sqrt{\pi} \varrho)^{D-3}}$$
with
$$\varrho= \sqrt{\gamma^2 (z-vt)^2+r^2}, \qquad \qquad r^2=  \delta_{ij} x^i x^j\,.  $$
the line element (\ref{STiso}) reads
\begin{equation}\label{STiso2}
\mathrm{d}s^2  = \left(\rule{0pt}{0.9em} 1+A  \kn \right)^{\!\raisebox{2pt}{$\scriptstyle\frac{4}{D-3}$}} \kn  \mathrm{d} {s}_{\mathcal{M}}^2  +  \biggr[\!\left(\frac{1-A}{1+A}\right)^{\!\akn 2}
- \left(\rule{0pt}{0.9em} 1+A  \kn \right)^{\!\raisebox{2pt}{$\scriptstyle\frac{4}{D-3}$}}
 \biggl] \gamma^2  ( \mathrm{d} {t}-v\kn \mathrm{d}{z})^{2},
\end{equation}
where $ \mathrm{d} {s}_{\mathcal{M}}^2= \eta_{MN}\kn \mathrm{d}x^{M} \mathrm{d}x^{N} $ is Minkowskian metric.

Thereby
$$ A = \frac{2 \kn G \Eps\, \Gamma\!\left(\frac{D-1}{2}\right)}{(D-2) (\sqrt{\pi} \varrho)^{D-3}  \gamma } \ll 1 $$
is \emph{sufficient} for validity of a Taylor expansion: retaining the first order, one obtains
\begin{equation}\label{STiso3}
\mathrm{d}s^2  =\left[1+ \frac{4 A}{D-3} \right]   \mathrm{d} {s}_{\mathcal{M}}^2  -\frac{4 \kn (D-2)\kn  A}{D-3}\, \gamma^2  ( \mathrm{d} {t}-v\kn \mathrm{d}{z})^{2},
\end{equation}

The non-zero components of this metric are:
\begin{align} \label{nozi}
&g_{00}=1-\frac{4   A}{D-3} \bigl[ (D-2)\kn\gamma^2-1\bigr]&& g_{0z}= \frac{4 \kn (D-2)\kn  A}{D-3}\, \gamma^2 v  \nn \\
&g_{zz}=-1-\frac{4     A}{D-3} \bigl[(D-2)\kn \gamma^2v^2+1 \bigr] && g_{x_{k}x_{k}}=-1-\frac{4    A}{D-3}
 \end{align}
which after substitution of $\vak^2 =16 \pi G$ exactly correspond
to those ones of the linearized solution $\bar{h}_{MN}$ (\ref{hpart})
\begin{align} \label{hpart123}
 \bar{h}_{MN}(x)=-\frac{\vak m\kn \Gamma\!\left(\fr{D-3}{2}\right)}{4\pi^{\frac{D-1}{2}}  \varrho^{D-3} }
 \left(u_M
u_N-\frac{1}{D-2}\,\eta_{MN}\right).
 \end{align}
Thereby the cut-offs due the moving particle are
\begin{align} \label{co_app1}
r_{\min} \simeq r_g\,,  \qquad\qquad z_{\min} \simeq r_g /\gamma\,.
 \end{align}
Introducing the energy-associated gravitational radius $\re$ as
$$\re =\frac{1}{\sqrt{\pi}}\biggl[\frac{2   \,
\Gamma\!\left(\frac{D-1}{2}\right)G \mathcal{E}}{D-2}\biggr]^{\frac{1}{D-3}}\!\!,$$
which is assumed to be constant independently of $\gamma$, one observes that when tending to the massless limit, the minimal-length cutoffs  due to the moving particle vanish as $r_{\min} \sim \gamma^{-1/(D-3)}$, $z_{\min} \sim \gamma^{-(D-2)/(D-3)}$.

If we live on the brane, piercing by the bulk black hole will excite explosive branons
whose energy can be transformed to matter on the brane. To calculate such effects on has, however, to apply different techniques.


\end{document}